\documentclass[aps,pre,reprint]{revtex4-1}

\usepackage{graphicx}  
\usepackage{booktabs}
\usepackage{amsfonts}
\usepackage{amsmath}
\usepackage{amssymb}
\usepackage{multirow}
\usepackage{color}
\usepackage{soul}
\usepackage{siunitx}

\begin{document}
\thispagestyle{empty}

\title{Computation of the Hydrodynamic Radius of Charged Nanoparticles from Non-equilibrium Molecular Dynamics}

\author{Lisa B. Weiss$^{a,b}$,Vincent Dahirel$^b$,Virginie Marry$^b$,Marie Jardat$^b$}
\affiliation{$^a$Faculty of Physics, University of Vienna, Boltzmanngasse 5, A-1090 Vienna, Austria, \\
$^b$ Sorbonne Universit\'e, CNRS, Physico-chimie des \'electrolytes et nano-syst\`emes interfaciaux, PHENIX, F-75005 Paris, France}

\date{\today}

\begin{abstract} 
We have used non-equilibrium molecular dynamics to simulate the flow of water molecules around a charged nanoparticle described at the atomic scale. These non-equilibrium simulations allowed us to compute the friction coefficient of the nanoparticle and then to deduce its hydrodynamic radius. 
We have compared two different strategies to thermostat the simulation box, 
since the low symmetry of the flow field renders the control of temperature non trivial.  We show that both lead to an adequate control of the temperature of the system. 
To deduce the hydrodynamic radius of the nanoparticle we have employed a partial thermostat, which exploits the cylindrical symmetry of the flow field. Thereby, only a part of the simulation box far from the nanoparticle is thermostated. We have taken into account the finite concentration of the nanoparticle by using the result of Hasimoto (J. Fluid. Mech. {\bf 1959}, {\it 5}, 317-328 ) for the friction force in a periodic cubic array of spheres. We have focused on the case of polyoxometalate ions, which are inorganic charged nanoparticles. It appears that, for a given structure of the nanoparticle at the atomic level, the hydrodynamic radius significantly increases with the nanoparticle's charge, a phenomenon that had not been quantified so far using molecular dynamics. The presence of an added salt only slightly modifies the hydrodynamic radius.

\end{abstract}  
\maketitle

  \section{Introduction}
  \label{introduction}
Charged nanoparticles are used in numerous technologies. Modeling the properties of dispersions of nanoparticles plays an important role to design applications\cite{Stark2015}. 
The numerical simulation of their properties usually requires the use of coarse-grained models, especially when nanoparticles are dispersed in aqueous solutions with added salt. Effective descriptions of the system averaged over water molecules and ions are justified, since the time scale of the nanoparticles transport is frequently much larger than typical solvent time scales. However, the definition of the parameters of coarse-grained models is often a complicated task.
Indeed, in order to perform simulations of the dynamical properties of nanoparticles in solution, one needs to determine their effective size and the behavior of water molecules in the vicinity of their surface. Both properties control the hydrodynamic radius of the nanoparticle. In most mesoscopic simulations techniques, like Brownian Dynamics (BD)\cite{ErmakJCP78,dahirel_JCP_2009}, Multi-Particle Collision Dynamics (MPCD) \cite{Padding2006,gompper_multi-particle_2008} or Dissipative Particle Dynamics (DPD) \cite{espanol2017}, the
boundary condition for the velocity of the solvent at the nanoparticles surface can be modeled as a slip or a stick boundary condition.
For colloidal particles, the definition of the hydrodynamic radius and of the boundary condition are relatively easy, because the structural radius, deduced for instance from microscopy or other structural investigations often matches the one deduced from the measurement of the diffusion coefficient using the Stokes-Einstein law (see for instance \cite{Kops1982}). 
For nanoparticles, whose size is not large compared to the solvent size, the whole definition of the hydrodynamic radius is not obvious anymore.
Does the solvent perfectly stick on the surface, or partially slip \cite{SchmidtJCP2003, SchmidtJPCB2004,BerneJPCB2012,IshiiPRE2016} ? If it sticks, is there an immobile layer of molecules at the surface of the nanoparticle that modifies the effective hydrodynamic size of the particle ? Does the charge influence the friction on the nanoparticle ? 
To answer these questions, modeling the system at the atomic scale is mandatory. For simple ions, the interpretation of individual transport coefficients in term of an effective radius raises many questions \cite{pauJPC90}.
It is known that the 
effective hydrodynamic radius depends on all interactions with the solvent molecules, and in particular that electrostatic forces give rise to a dielectric contribution to the friction \cite{Hubbard,Wolynes80}.   In the present study, our goal is to compute the flow of water molecules around a charged nanoparticle from non-equilibrium (NEQ) atomistic simulations, in order to derive quantities like 
the hydrodynamic radius that can be used afterwards in coarse-grained simulations, and that can be used to challenge typical electrolyte or colloidal concepts and theories for small nanoparticles.
Such non-equilibrium molecular dynamics simulations have to be carried out and analysed with great care.

A frequently employed strategy in molecular dynamics to create a solvent flow is to apply an external force to the system, 
either on all solvent atoms, like in Poiseuille flow, or  only on charged particles in the solvent as in the case of electroosmosis \cite{BarratJCP2014,AluruJCP2017A}. 
In both cases, some energy is added to the system, and is converted into viscous energy. It results in heating of the system. Thus, a thermostat must be used in order to extract the added energy and to ensure a constant temperature during the simulation.
Simulations of water flows between parallel walls, or parallel layers of inorganic materials modeled at the atomic level are well described in the literature \cite{JolyJCP2006,Botan2011,botan_how_2013,Bonthuis2013}. Several studies have focused on the different thermostat strategies for the simulation of fluids confined in a channel \cite{Bernardi2010,Yong2013, DeLuca2014}. In such cases, 
one can take advantage either of the high symmetry of the solvent flow or of the presence of the numerous  atoms of the walls to thermostat the system. For example, an efficient strategy consists in coupling only the wall atoms with the thermostat \cite{DeLuca2014}.
However, when only one fixed obstacle is put in the center of a simulation box filled with solvent molecules, the solvent velocity field varies in all directions and the use of usual thermostat strategies is tricky.  We focus here on two different procedures to thermostat the simulation box: The Profile Unbiased Thermostat\cite{Evans1984a} (PUT) and the Partial Thermostat (PT).
The principle of the PUT is the following. The simulation box is divided in cells,  and velocities of solvent molecules relative to the average solvent velocity in this cell are rescaled so that the target temperature is obtained, via a Nos\'e-Hoover algorithm \cite{NoseHoover}. 
When the symmetry of the velocity field is  low, using PUT demands a particular study of the choice of the size of the cells. 
If they are too large, the flow field varies too much inside the cells. If they are too small, the number of degrees of freedom to be thermostatted together is too small. The other procedure, called in what follows the Partial Thermostat (PT) is simpler in principle: Only a fraction of the simulation box is thermostatted, and only part of the degrees of freedom of water molecules are thermostatted. This is similar to the aforementioned thermostatting of the atoms of a wall \cite{DeLuca2014}. The challenge in this case is to define a region where the geometry of the velocity field is rather uniform, and to check that this region is large enough to thermostat the rest of the simulation box.

In this paper, we present a robust way to study the flow around a fixed, approximately spherical nanoparticle from non-equilibrium molecular dynamics. We thoroughly studied the efficiency of the PUT and of the PT in this case, and defined the conditions where they lead to a proper control of the temperature of the system. From these NEQ simulations,  we computed the friction coefficient exerted by the solvent on the nanoparticle. Then, we deduced the hydrodynamic radius of the nanoparticle by using the approximate analytical result of Hasimoto \cite{Hasimoto,Yeh2004}, which gives the flow field across an infinite periodic array of spheres.  
 Our simulation strategy is applicable to any nanoparticle that can be approximated by a sphere. Nevertheless, as water-interface interactions are highly dependent on subtle surface details, we focused here on a specific nanoparticle. Indeed, even charged generic nanoparticles can have spurious hydrophobic properties \cite{Dzubiella2004}. We focused on polyoxometalate ions (POM), which are inorganic charged nanoparticles used as  a standard in electrokinetic measurements, especially in electroacoustic experiments \cite{Hodne2001}. More precisely, we took the example of the phosphotungstate  anion [PW$_{12}$O$_{40}$]$^{3-}$. 
 This family of systems is already well studied in the literature, but a description of their dynamic characteristics at the atomic scale is still lacking.  
Once the simulation procedure has been validated, we could make a preliminary study of the influence of the charge of the nanoparticle on its dynamic properties. We computed the friction coefficient and the hydrodynamic radius of a neutralized POM, and of nanoparticles of charge $-3 e$ $-4 e$, and $-5 e$ while keeping the surface charges unchanged. We also studied the influence of the presence of an added salt on the hydrodynamic radius of these nanoparticles. 
  
The paper is organized as follows. Section II describes the force fields and the procedures used to impose a flow and to compute the friction coefficient. In Section III we compare the efficiency of the profile unbiased thermostat (PUT) and that of the partial thermostat procedure. The calculation of the friction coefficient of the inorganic ion is discussed in Section IV. 

\section{Methods}
\label{methods}
\subsection{Modeling of the system}

\begin{figure}
\begin{center}
\includegraphics*[scale=0.4]{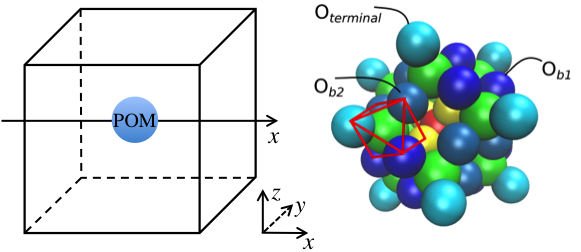}
\end{center}
\caption{{Left: sketch of the simulation box, containing a POM ion in the center of the box. The flow is applied along the $x$-axis. Right: Structure of the POM ion ($\rm [PW_{12}O_{40}]^{3-}$). The central atom P is in red; Surrounding O atoms forming a tetrahedron are in yellow; Surface O are in blue,  and the W atom is in green. Each W atom is surrounded by an octahedron of O atoms. }}
\label{sketch}
\end{figure}

As stated in the introduction, we are interested in the flow of water molecules around a specific polyoxometalate ion (POM): the phosphotungstate  anion [PW$_{12}$O$_{40}$]$^{3-}$. Counterions are potassium ions $\rm K^+$, and in some cases chloride ions $ \mathrm{Cl}^{-} $ have also been added. 
POMs are inorganic anions consisting of a central atom surrounded by MO$_n$ polyhedrons, where M is a metal atom and O oxygen. The most known class of polyoxometalates is the Keggin anion that has the general formula [XM$_{12}$O$_{40}$]$^{n-}$, depicted in the right part of     Fig. \ref{sketch}. In [PW$_{12}$O$_{40}$]$^{3-}$, the central phosphorus is surrounded by oxygens forming a tetrahedron. Each of these oxygen atoms forms the corner of an octahedron, where all corners are formed by oxygen atoms, which enclose the tungsten. Therefore, the surface atoms accessible to the solvent are oxygen atoms shown in various blue shades on    Fig. \ref{sketch}, which indicate distinct distances to the central atom.
The POM is placed in the center of a  cubic simulation box  as depicted in the left part of    Fig. \ref{sketch}. Periodic boundary conditions are applied in all three directions. The box is filled with water molecules. In NEQ simulations, an external force is applied on water molecules along the $x$-axis.

Several previous theoretical investigations of polyoxometalates focused on their electronic properties using 
density functional theory (DFT). 
The optimized DFT structures 
for the Keggin anions [PW$_{12}$O$_{40}$]$^{3-}$ and [SiW$_{12}$O$_{40}$]$^{4-}$ 
well coincide with experimental data as cited in reference \cite{Maestre2001}. 
Some of these authors developed force fields for Keggin anions containing phosphorus, 
silicon and aluminium as central atom \cite{Lopez2005b,Leroy2008b}. 
The proposed force fields were successfully engaged to analyse ion pair formation between these POMs and potassium, sodium, and lithium as counterions. 
For a box size with $1000$ water molecules and one POM compared to $8000$ water molecules and $8$ POMs no size dependence of the radial distribution functions was noticed\cite{Leroy2008b}.
These force fields were also successfully employed to study the aggregation of multiples POMs as function of counterions, charge and solvent \cite{Chaumont2008b,Chaumont2013b}. 
Therefore, we have taken the force field parameters from Ref. \cite{Leroy2008b}. The parameters of the force field are given in Table \ref{POM-SIM}. As the coordinates of the DFT optimized POM structure used in Ref. \cite{Leroy2008b} are not available, 
the crystal coordinates of H$_3$PW$_{12}$O$_{40}\cdot6$H$_2$O were taken from reference \cite{Noespirlet1975}. 
We checked that the intramolecular radial distribution functions are in very good agreement with those reported
in Ref. \cite{Leroy2008b}.  The extended simple point charge model of water (SPC/E) was employed \cite{Berendsen1987}.
The force field parameters for potassium and chloride ions were taken from references \cite{Lee,Koneshan1998}. 
 Lorenz-Berthelot rules were used 
to calculate the interatomic interactions for different atom types.

 \begin{table}
 \caption{{Force field parameters for the [PW$_{12}$O$_{40}$]$^{-3}$ ion and distances of the different atoms to the center of the POM.}}
 \centering
 \begin{tabular}{|c|c|c|c|c|}
 \hline
 $i$   & $d_{P-i}\, [\si{\angstrom}]$&  $\epsilon_{i-i}$ [$kcal.mol^{-1}$]&$\sigma_{i-i}\,[\si{\angstrom}]$&$q$ [$e$]\\ 
 \hline
  P&0.00& 0.2453&3.0&1.51\\
  W&3.56&0.2211&2.34&3.81\\
  O$_{terminal}$&5.24&0.2145&3.17&-0.85\\
  O$_{b1}$&3.93&0.2145&3.17&-1.37\\
  O$_{b2}$&3.37&0.2145&3.17&-1.55\\
  O$_{tetra}$&1.57&0.2145&3.17&-1.2475\\
  \hline
 \end{tabular}
 \label{POM-SIM}
 \end{table}

 All simulations were performed using the molecular dynamics simulation package LAMMPS \cite{lammps}.
The SHAKE algorithm, which is implemented in LAMMPS as described in reference \cite{Ryckaert1977} was used to restrict bond lengths and angles of the water molecules.  
Moreover, the POM was kept fixed in the center of the box during the equilibrium and non-equilibrium simulations. 
The forces acting on all the atoms of the POM were thus set to zero. 

Long range coulomb interactions were taken into account via a particle-particle particle mesh solver.
A cut-off radius of  $\SI{15.0}{\angstrom}$ was set for the Lennard-Jones interactions and the real space Coulomb interactions. 
For each initial configuration of the system,  a combination of equilibration runs in the $NVT$ and $NpT$ ensembles were performed to obtain bulk water at the expected density, for approximately $\SI{1}{\nano\second}$, with a time step $\delta t=1.0\,\si{\femto\second}$.     The target values were  $T=\SI{300}{\kelvin}$ and $p=\SI{1.0125}{\mega\pascal}$.
 During equilibration, pressure and temperature control were reached by a Nos\'e Hoover barostat and thermostat. 
These are implemented in LAMMPS as described in reference \cite{Martyna1994,Martyna1992}. No thermostat was used for ions. We verified that ions were effectively thermostatted by the surrounding solvent.
 
To induce a flow of water molecules in the simulation box,  two strategies are described in the literature: Either adding a force to the fluid particles (within a slice or for all 
fluid molecules), or 
imposing a constant velocity to part of the atoms of the fluid. We have applied a constant force to all atoms in the $x$-direction, which is equivalent to a pressure driven flow. We have varied the force until 
the flow corresponds to a low Reynolds number $R_e\ll1$, defined as:
\begin{align}
R_e=\frac{v_{H_2O\times d_{POM}}}{\nu}
\end{align}
with a diameter of the POM $d_{POM}\approx\SI{12}{\angstrom}$ 
and the kinematic viscosity denoted by $\nu$. We have then ensured that the flow is in the linear response regime. The characteristic velocity of the fluid $v_{H_2O}$ can be defined as $v_{H_2O}= \frac{|{\langle F_x\rangle }|}{\xi^S}$, where $|{\langle F_x\rangle }|$ is the total external force acting on solvent molecules along the $x$-axis, and is therefore an input of the simulation, and $\xi^S$ is the friction felt by a perfectly spherical obstacle of radius $R$ with stick boundary conditions computed by the Stokes law $\xi^S=6\pi \eta R$. 
If we take the viscosity of SPC/E water ($\eta_{water}=0.68\times10^{-3}$~Pa.s at 300~K, see reference  \cite{TaziJPCM2012}), and a force equal to $f=0.1\times10^{-13}$~N per water molecule ($N\simeq1700$ water molecule in the simulation box), we obtain a Reynolds number $R_e\simeq0.003$. 

During the non-equilibrium simulations, the central nanoparticle is fixed in the center of the simulation box and is not allowed to rotate. We have checked that the velocity profiles of water molecules are the same whatever the orientation of the nanoparticle within the statistical error. 

\subsection{Use of a thermostat}
Our first goal is to establish a strategy to thermostat a flow around an approximately spherical obstacle in molecular dynamics simulations. Hence, we simulate a system out of equilibrium, but still in the linear response regime. As already mentioned, such   
non-equilibrium molecular dynamics  requires special thermostatting strategies. 
In this study, NEQ simulations were performed employing two different thermostatting strategies, namely the profile unbiased thermostat (PUT)\cite{Evans1984a} and a partial thermostat (PT).

\subsubsection{Profile Unbiased Thermostat}
\label{methodPUT}
The PUT has been developed by Evans \textsl{et al.}\cite{Evans1984a}  to thermostat a turbulent flow of Lennard-Jones particles. 
In this procedure, the simulation box is divided into a fixed number of bins. Subsequently, the center of mass velocity of every bin is calculated. 
The velocities of the solvent particles minus the center of mass velocity are computed, 
and these velocities are employed for temperature definition \cite{Evans1984a}. More precisely, 
for every bin of position ${\mathbf{r}}$ and at each time step $t$, 
the local kinetic energy  $E_{kin}(\mathbf{r},t)$ is calculated by summing the square of all velocities $\mathbf{v}_i$ minus 
the average streaming velocity $\mathbf{u}$ over all atoms $N$ in the bin, which allows one to compute a local and instantaneous temperature $T\left(\mathbf{r},t\right)$: 
\begin{align}
E_{kin}(\mathbf{r},t)&=(D\cdot n(\mathbf{r},t)-d)\frac{k_BT(\mathbf{r},t)}{2} \\
&=\sum_{i=1}^{N}\frac{1}{2}m_i[\mathbf{v}_i(\mathbf{r},t)-\mathbf{u}(\mathbf{r},t)]^2\delta(\mathbf{r}_i(t)-\mathbf{r})
\label{equ-put}
\end{align}
Here the number of degrees of freedom per particle $D$ is multiplied by the instantaneous 
number of molecules $n(\mathbf{r},t)$ in the bin. The dimensionality $d$ of the system is subtracted to account for the conservation of 
the $d$ components of the momentum of the simulation box; $k_B$ is Boltzmann constant. 
The number of degrees of freedom per particle is $D=6$ for the rigid SPC/E water model 
in combination with the SHAKE algorithm. The delta function $\delta(\mathbf{r})$ ensures that all atoms are within the specific bin. The system temperature is then calculated, and the Nos\'e-Hoover thermostat with a relaxation time $\tau$=1000 fs is used. 
We have checked that our results were not influenced by the value of the relaxation time: a value twice as  large leads to the same results within the statistical error.
In the PUT procedure, the number of bins in the simulation box is a crucial parameter:
The resolution of the grid should be high enough 
to have a small velocity gradient within every bin, but small enough to ensure  
that thermal fluctuations are statistically significant inside each bin. This common idea is a classical result for random variables, as quantified for instance by Chebyshev's inequality for the random variable $[\mathbf{v}_i(\mathbf{r},t)-\mathbf{u}(\mathbf{r},t)]^2$. 
To illustrate this, one can  
simplify the above equation writing for a fluid at rest and particles of equal masses $m_i$ : $$T(\mathbf{r},t)=\frac{A}{N}\sum_{i=1}^{N}\mathbf{v}_i(\mathbf{r},t)^2\delta(\mathbf{r}_i(t)-\mathbf{r})$$ 
where A does not depend on $N$. 
Chebyshev's inequality then reads:
\begin{align}
P(|T(\mathbf{r},t)-T|>\delta v^2)&<\frac{1}{N}
\label{bienayme}
\end{align}
where $P$ is a probability, $T$ is the temperature of the system ({\it i.e} the temperature computed for a large number of degrees of freedom), and $\delta v^2$ the standard deviation of the individual square 
velocities $\mathbf{v}_i(\mathbf{r},t)$.
$T(\mathbf{r},t)$ converges towards a good measure of the exact temperature $T$ for large values of $N$ (large bins), for a fluid at rest.  

\subsubsection{Partial Thermostat}
\label{methodPT}
Another way to regulate the temperature of the simulation box during NEQ simulations consists in using a partial thermostat. 
Typically, the simulation box is divided in two parts: The first one is thermostatted by using a NH algorithm and the second one is not thermostatted. It means that molecules located in the second part of the simulation box would evolve in a ($NVE$) ensemble if no heat transfer were allowed with the first part of the box. 
This strategy has been used for simulating a flow in a porous media, for which the atoms of the solid walls are thermostatted, but the atoms of the fluid are not \cite{DeLuca2014}.
In our case, we chose to thermostat the fluid in a region where the averaged fluid velocity is very weak in the $x$ and $y$ directions. 
Therefore, the thermostatted region should be as far as possible from the spherical obstacle, and sufficiently large to contain a significant number of fluid molecules.
More precisely, the simulation box has been separated in two parts: (1) The molecules that are within a cylinder of radius $R_{cyl}$, the axis of which  is parallel to the $x$-axis passing through the central atom of the POM and the length of which is equal to that of simulation box, 
and (2) the water molecules that are outside the cylinder. 
The molecules that are outside the cylinder are thermostatted by a Nos\'e Hoover (NH) thermostat restricted to 
the two components of the velocity perpendicular to the flow ($v_y$ and $v_z$).
For such a thermostat ($T_{yz}$) the number of degrees of freedom is $4$ for rigid water molecules, so that
 the total number of degrees of freedom to be accounted for by the thermostat is 
$4\,n_{molec,out}-2$. This expression takes into account the fact that the overall transport in $y$ and $z$ direction should be zero due to the geometry of the flow profile in the thermostatted region as soon as $R_{cyl}$ is high enough. 

\subsection{Computation of the friction coefficient of the solute}
\label{methodfriction}
The friction coefficient $\xi$ of a solute can be defined as the ratio of the force exerted by the fluid on the solute to the characteristic velocity of the fluid.  
This force is indeed proportional to the fluid velocity in the linear regime, under the 
conditions of a low Reynolds number. 
 The friction on a sphere is unambiguously defined for a sphere immersed in a fluid at infinite dilution, 
when the fluid is moving along the $x$-direction far from the object with a velocity $v_{x,\infty}$:
\begin{align}
\xi=\frac{|{\langle F_x\rangle }|}{v_{x,\infty}}
\label{frictioninf}
\end{align}
In  eq. \ref{frictioninf}, $\langle{F_x}\rangle$ is the total force acting on the sphere, averaged over the molecular dynamics trajectory. In a low Reynolds number flow, for stick boundary conditions, the classical Stokes calculation yields $\xi=6 \pi \eta R$, where $R$ is by definition the hydrodynamic radius of the spherical obstacle \cite{Guyon2001}.

In our case, there is one solute fixed in the center of the simulation box with periodic boundary conditions, as it is commonly done in molecular dynamics, so that the solute concentration is finite. None of the fluid molecules  inside the simulation box can be considered to be far from
the solute.
However, the long-range nature of the hydrodynamic interactions influences dynamical quantities such as the Stokes friction, even when structural equilibrium properties are close to those of an infinite fluid around a unique spherical object \cite{kremer93}. 
 We are actually dealing with a flow of solvent 
 in a periodic cubic array of spherical objects. Since the fluid is incompressible,
 the total mass flow across the $(y,z)$ planes perpendicular to the flow is conserved for all values of $x$. In the $(y,z)$ planes that do not cross the spherical obstacle, the average fluid velocity $v_{av}$ does not depend on the value of $x$. The friction coefficient $\xi$ can then be unambiguously defined as the ratio between the averaged fluid velocity $v_{av}$ and the total applied force $|{\langle F_x \rangle }|$   : 
\begin{align}
\xi=\frac{|{\langle F_x \rangle }|}{v_{av}}
\label{friction}
\end{align}

In this case, the friction coefficient $\xi$ in the Stokes regime does not scale as the hydrodynamic radius. Corrections related to the size of the particle and to the distance between the periodic images (here the box length $L$) must be taken into account.
This hydrodynamic problem has been studied in various works, and the impact of periodic images on friction has been first estimated by Hasimoto \cite{Hasimoto}. In the
case of a periodic cubic array of spheres of same radius $R$, with $R$ being very small compared
to the distances between spheres $R\ll L$, one can neglect quadratic corrections and gets the following expression of the force acting on a sphere:

\begin{align}
\mathbf{F}=\frac{6\pi\eta R}{1+2.8373 R/L} \mathbf{v}_{av}
\label{frictioncorrection}
\end{align}

In the regime where $R$ is not very small compared to $L$, Hasimoto derives a more precise analytical expression of the force (see Section 5 of Hasimoto's paper \cite{Hasimoto}). In what follows, we have used this refined expression to deduce the hydrodynamic radius of the charged nanoparticle from our non-equilibrium simulations. Indeed, as we perform explicit solvent calculations with many solvent molecules, we are constrained in a regime where the size of the simulation box is not much higher than the size $R$ of the nanoparticle. Then, the linear correction in $R/L_{box}$ is not accurate, and one needs to consider higher order terms.  

To compute the friction coefficient from the simulations, we have to evaluate the total force acting on the nanoparticle. When the flow is stationary, the force acting on the nanoparticle is exactly equal to the total external force applied to solvent molecules. Indeed, when the fluid does not accelerate, the total sum of all forces on the fluid should be zero. The forces acting between water molecules, between ions, and between ions and water molecules cancel out due to Newton's third principle. The remaining contributions are thus the forces acting on the solvent due to the presence of the central nanoparticle, and the external forces. These two forces are then opposite. The force felt by the nanoparticle should therefore be exactly equal to the external force. This result was numerically confirmed in our simulations. As we proceed to show in Sec. \ref{efficiency-therm}, the chosen thermostat strategy has no influence on the balance of forces.. In practice, we have first computed the friction coefficient as the ratio between the total force acting on the nanoparticle and the average fluid velocity $v_{av}$. We have then deduced the hydrodynamic radius of the nanoparticle by using Hasimoto's expression given in Section 5 of reference  \cite{Hasimoto}.

In the following, we define a characteristic friction $\xi^o$ as the friction felt by a perfectly spherical obstacle of radius $0.6~$nm in a periodic cubic array of period $3.7$~nm, which is the value of the simulation box length in our calculations. 
With a viscosity equal to $ 0.68\times 10^{-3}$ Pa.s, which is the viscosity at 300~K of pure SPC/E water \cite{TaziJPCM2012},
we obtain the value $\xi^o=1.381$ $10^{-11}$ kg.s$^{-1}$ using Hasimoto's expression given in Section 5 of reference  \cite{Hasimoto}. 
For the same parameters, we define as a reference the fluid velocity $v^o= \frac{|{\langle F_x\rangle }|}{\xi^o}$, where $|{\langle F_x\rangle }|$ is the total external force acting on solvent molecules in the $x$ direction. In what follows, there are about $1700$ water molecules around a POM in the simulation box, with an external force on each water molecule of $0.1\times10^{-13}$ N: $v^o=12.31$ m.s$^{-1}$.

\section{Results}
\label{results}
\subsection{Efficiency of the thermostat to compute the flow around a nanoparticle}
\label{efficiency-therm}
\subsubsection{Profile Unbiased Thermostat}
In this part, we test the ability of the Profile Unbiased Thermostat to regulate temperature so that the flow of solvent in non-equilibrium molecular dynamics is correct. The flow depends on the temperature for various reasons. The most obvious one is that the fluid viscosity highly depends on temperature.  
The effect of the thermostat on the fluid viscosity is thus indirectly evaluated through the measure of the average velocity $v_{av}$  in
the $(y,z)$ planes perpendicular to the flow. Indeed, for a given external force, this average velocity is inversely proportional to the friction coefficient $\xi$, which is itself exactly proportional to the viscosity. Therefore, in a Stokes flow, the viscosity is inversely proportional to the average velocity $v_{av}$.  

As already mentioned, the correct use of the PUT requires to choose the number of bins used to compute the local temperature carefully.
In this section, we report the results we obtained for a flow of water around a POM nanoparticle at small Reynolds numbers, 
using the PUT procedure with several resolutions of the grid of bins. 
In this case, as the POM is described at the atomic scale, the velocity field of water is highly inhomogeneous in the vicinity of the POM. 
In principle, we should use small bins in the PUT to account for spatial variations of the velocity field. 
But, there should be enough molecules per bin so that thermal fluctuations inside each bin converge close 
to the average fluctuations in a bulk fluid at the correct temperature. We therefore need to evaluate the effect of the bin size on the flow for reasonable values of the bin size.

As far as the flow inhomogeneity is concerned, choosing bins with a size of half the box length $L_{box}$, {\it i. e.} taking a grid of $2^3$ bins, does not improve the problem, because of obvious symmetry reasons. 
We thus chose bins of size equal to  $L_{box}/4$ as the largest possible bins.   
In our case, there are about $1700$ water molecules in a cubic box of about $37$~\AA~length, so that a grid of $10^3$ bins corresponds to  about $2$ water molecules per bin in average. For $8^3$ bins there are about $3$ molecules per bins, 
$8$ for $6^3$ bins, and $27$ molecules for $4^3$ bins. We have chosen as greatest resolution, {\it i. e.} as the smallest bins, a case where  the average number of water molecules is greater than $2$, else the calculation makes no sense at all. 
Finally, we have thus restricted our study to the cases with $8^3$,  $6^3$, and  $4^3$ bins. 
In all described simulations a force of
$f=0.1\times10^{-13}$ N was added to all water molecules.

\begin{figure}
\begin{center}
\includegraphics*[scale=0.3]{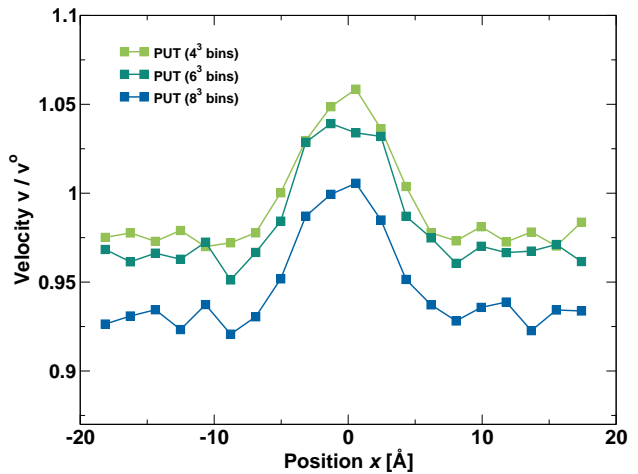}
\end{center}
\caption{{Velocity profiles of water molecules along the $x$-direction for different resolutions of the PUT grid. The reference velocity $v^o$ is the average velocity around a perfectly spherical obstacle periodically fixed within a cubic array \cite{Hasimoto}. The center of the POM is located at $x=0$. }}
\label{velocity-prof-PUT-bin}
\end{figure}

We give in    Fig. \ref{velocity-prof-PUT-bin} the velocity profiles of water molecules as a function of the $x$ coordinate, for different resolutions of the PUT grid. The results weakly depend on the grid resolution. 
The standard uncertainty of the average velocity has been computed from a block analysis of the fluctuations of the velocity of the fluid over several temporal averages of $4$ ns. It is always smaller than $0.35$~m.s$^{-1}$ so that the error estimate on $v/v^o$ is equal to $0.03$.
When the grid size changes from $4^3$ to $6^3$ bins, the differences in the averaged flow velocity  $v_{av}$ are smaller than these fluctuations.  
There is no exact analytical result with which we could directly compare the values of the velocity  $v_{av}$. 
Nevertheless, if the POM nanoparticle were a perfectly spherical particle of radius equal to $6.0$~\AA ~with stick boundary conditions, 
one would get $v_{av}=v^o$, where $v^o$ is computed for the Stokes flow within a cubic array of spherical obstacles, 
with spheres of radius $6.0$~\AA~ close to the radius of POM. 
The results are close to this perfectly spherical case. 
The velocities obtained with thermostat grids of $8^3$ bins are nevertheless smaller than the velocities obtained with larger bins. 
This difference is slightly larger than the uncertainty of the velocity. It can be explained by the small size of the bins. 
As indicated before, with such a small number of water molecules in each bin (about 3), the thermal fluctuations might not be significant from the statistical point of view.

There are then two indications showing that the PUT is adequate to regulate the temperature of the fluid. Nevertheless, using a Nos\'e Hoover algorithm close to the nanoparticle might change the local dynamics, such as for instance the local diffusional dynamics of water and ions. If one is interested in the surface conductivity or other surface phenomena, it is useful to see if another thermostatting strategy can be used which does not affect the area close to the solid-liquid interface. 

\subsubsection{Partial thermostat with a cylindrical geometry}

We use in this part a partial thermostat with a cylindrical geometry. In
parts of the simulation box far from the obstacle, outside of the cylinder, the flow field is mainly oriented towards the direction of the applied force. 
This subpart is thermostatted in the directions perpendicular to the flow.
This requires to adequately choose the radius of the cylinder. 
As it can be seen on    Fig. \ref{velocity-prof-PUT-bin}, 
for $x$ coordinates between $-18$~\AA~and $-10$~\AA~the velocity of the fluid in the $x$ direction is almost constant, with variations below the uncertainty of our calculations (the center of the obstacle is in the position $x=0$). 

The tendency is the same for all regions of space further than 
$15.0$~\AA~
 from the center of the obstacle, and this will be checked {\it a posteriori}.   
We choose thus to thermostat the solvent outside a cylinder of radius equal to $17.0$~\AA. 

\begin{figure}
\begin{center}
\includegraphics*[scale=0.3]{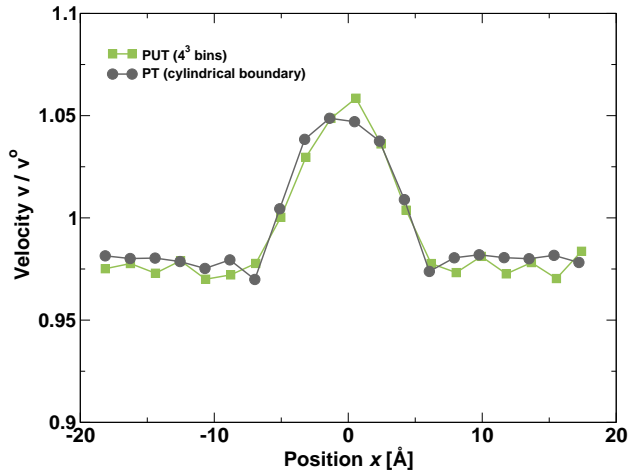}
\end{center}
\caption{{Velocity profiles of water molecules along the $x$-direction obtained with the partial thermostat with a cylindrical frontier and compared to those obtained with the PUT for a resolution of $4^3$ bins. The center of the POM is located at $x=0$.}}
\label{velocity-prof-PUT-PT}
\end{figure}

First, this cylindrical thermostat was used for equilibrium simulations, with a target temperature of $\SI{300}{\kelvin}$. A perfect agreement was obtained between the time averaged temperature 
within this procedure and that obtained with a NH thermostat on the  full simulation box.
We turned then to non-equilibrium simulations. 
A $\SI{2}{\nano\second}$ non-equilibrium run was done to reach steady state before the production run started. 
At steady state, we expect that the
molecules inside the cylinder keep a constant temperature through heat transfer
with the thermostatted region of the box. Indeed, 
no heating was observed over $\SI{20}{\nano\second}$: The mean temperature averaged over time and over the whole simulation box have been computed {\it a posteriori} using the simulation trajectory. It is $\SI{299.5}{\kelvin}$, and thus it agrees perfectly with the target temperature of the $T_{yz}$ thermostat. 
Moreover, the homogeneity and the time invariance of the temperature inside the simulation box have been checked: 
The temperature was computed in subregions of the simulation box, within hollow cylinders with the same axis as the
thermostat boundaries, over durations of $\SI{0.2}{\nano\second}$ along the production run. All cylinders keep a temperature of $\approx\SI{300}{\kelvin}$. 
As with PUT, the velocity profile averaged over successive durations of $4\,\si{\nano\second}$ fluctuates by $\SI{1}{\metre\per\second}$.
As shown in    Fig. \ref{velocity-prof-PUT-PT},
the averaged velocity of water molecules obtained when using the Partial Thermostat 
is very close to (i) the one obtained with the analytical formula for an array of spherical obstacles, i.e. it is close to $v^0$, and (ii) the one obtained with the Profile Unbiased Thermostat.
This agreement is a strong evidence showing the ability of both thermostats to regulate temperature in non-equilibrium simulations around a quasi-spherical obstacle.  

\begin{figure}
\includegraphics[scale=0.4]{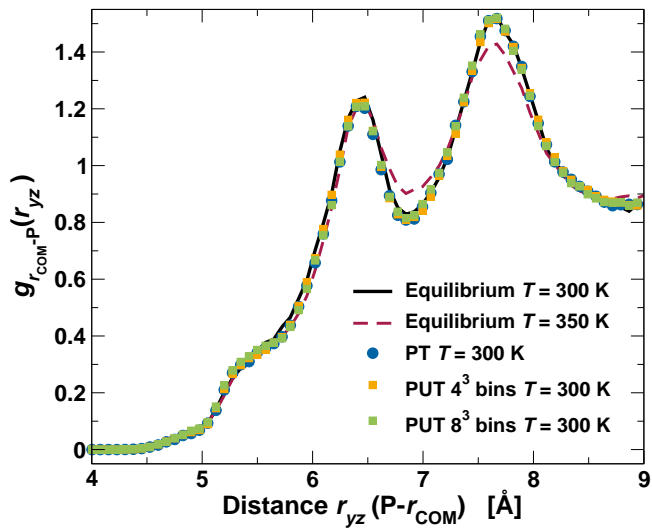}
\caption{Distribution functions between the center of mass of water molecules and the phosphorus atom of the POM computed in a cylindrical bin centered on the POM, of length $d_{bin}=\SI{0.373}{\angstrom}$, as a function of $r_{yz}=\sqrt{y^2+z^2}$. Comparison between results obtained: From equilibrium simulation at $\SI{300}{\kelvin}$ (black line), from equilibrium simulation at $\SI{350}{\kelvin}$ (red dotted line), 
and  from non-equilibrium simulations with a partial thermostat with a cylindrical frontier (blue circles) at $\SI{300}{\kelvin}$ and with a PUT with $4^3$ or $8^3$ bins. }
\label{cylinder-gr}
\end{figure}

Finally,
we have checked that the density distribution functions 
around the spherical obstacle correspond to that of an equilibrium simulation at the right temperature ($\SI{300}{\kelvin}$), as the
local temperature is expected to affect water structure.  
Equilibrium simulations for $T=\SI{350}{\kelvin}$ were also done.  
The significantly higher temperature of $\SI{350}{\kelvin}$ has been chosen to obtain a quantitative effect on distribution functions, one order of magnitude higher than noise. We compare the results obtained from these two sets of equilibrium simulations with 
non-equilibrium simulations using the Partial Thermostat and the Profile Unbiaised Thermostat.
We show on    Fig. \ref{cylinder-gr} the distribution functions between  hydrogen atoms of water and the phosphorus atom of the POM. They have been computed in a slice perpendicular to the $x$ direction with a width of  $d_{bin}=\SI{3.1}{\angstrom}$. Within the slice, the rdf is computed in cylindrical bins of length $d_{bin}=\SI{0.373}{\angstrom}$ centered on the POM, as a function of $r_{yz}=\sqrt{y^2+z^2}$. As expected, increasing temperature makes water molecules more mobile and less structured:
The maximum of the second peak of the distribution function  decreases strongly with increasing temperature. The distribution function obtained from non-equilibrium simulations coincides very well with that obtained at equilibrium at the same temperature. In particular, the difference at the second peak ($r_{yz}$ around $\SI{7.6}{\angstrom}$) is much weaker than the difference with the distribution function obtained at $\SI{350}{\kelvin}$. The agreement is observed whatever the size of the grid when the PUT is applied.
This also confirms that there is no significant heating of the system at the solid/water interface with the partial thermostat, even if the thermostat only works far from this interface.

In conclusion, we have identified two ways of properly thermostatting the simulation box in non-equilibrium simulations of liquid flow around a spherical obstacle: (i) The Profile Unbiased Thermostat, and (ii) the use of a cylindrical Partial Thermostat.
Both methods allow us to obtain the same velocity profiles of water molecules at steady state, and the same density distribution functions around the central nanoparticle.  Still, in principle the Nos\'e Hoover algorithm may affect the local diffusional dynamics of molecules even if the average properties (velocity profiles, distribution functions) are correct. The friction coefficient we are interested in  is especially influenced by the  structure and dynamics of water molecules close to the central nanoparticle.  In what follows, we have then only used the partial thermostat with a cylindrical symmetry. This thermostat only affects the components of water velocities in the direction perpendicular to the flow ($T_{yz}$), and far from the central nanoparticle. This also ensures that the balance of forces on the $x$-direction, which we use to compute the friction coefficient (see eq. \ref{friction}), is not influenced by the thermostat.

\subsection{Friction coefficient of charged inorganic nanoparticles }

Using the cylindrical partial thermostat, we have computed the friction coefficient of the POM and of inorganic nanoparticles analogous to the POM but with other charges, also in the presence of added salt, from non-equilibrium simulations. From this quantity, as described before, we have used Hasimoto's formula to deduce the hydrodynamic radius of the nanoparticle, assuming stick boundary conditions. 

Ten different systems were studied, which are listed in Table \ref{MDsystems}. The exact number of water molecules in the simulation box and the precise size of the box, obtained after equilibration runs in the $(NpT)$ and $(NVT)$ ensembles are also given. The counterion of the POM is 
potassium $K^+$ in every case.  In some cases, potassium chloride is added, with a total number of potassium ions always equal to $15$, so that the electrostatic screening is the same for all systems. This corresponds to a concentration of $c_{\mathrm{K}^+}=0.48\,\si{\mole\per\liter}\approx\SI{0.5}{\mole\per\liter}$. 
The number of chloride ions is adapted to ensure electroneutrality. 

\begin{table}
\centering
  \caption{{Simulated systems with abbreviations. The box length $L_{box}$ is given in $\si{\angstrom}$. Concentrations 
$c_{\mathrm{POM}}$, $c_{\mathrm{K}^+}$ and $c_{\mathrm{Cl^-}}$ are given  in mol.L$^{-1}$,  $q_P$ is the charge on the phosphorus in multiples of the elementary charge $e$.}}
  \begin{tabular}{|c|c|c|c|c|c|c|}
  \hline
   simulation & $L_{box}$&$n_{H_2O}$ &$c_{\mathrm{POM}}$&$c_{\mathrm{K}^+}$& $c_{\mathrm{Cl^-}}$&$q_P$\\

  \hline
  POM0  & 37.32456 & 1705 &  0.032 &0.0  &0.0 &4.51\\ 
  POM0-salt& 37.35643 & 1676& 0.032& 0.478&0.478&4.51\\
  POM3  & 37.21562 & 1702&   0.032& 0.097&0.0  &1.51\\
  POM3-salt& 37.42509 & 1681&  0.032& 0.475&0.380  &1.51\\
  POM4  & 37.35305 & 1701&   0.032& 0.127 &0.0  &0.51\\
  POM4-salt& 37.27600 & 1681&  0.032& 0.481&0.353  &0.51\\
  POM5  & 37.18164 & 1701&   0.032& 0.162&0.0  &-0.49\\
  POM5-salt& 37.27939 & 1682&  0.032& 0.481&0.321  &-0.49\\
  \hline
  \end{tabular}
      \label{MDsystems}
  \end{table}

To obtain nanoparticles of different charges, the charge of the central atom is increased or reduced compared to that given in  Table \ref{POM-SIM}. 
The goal is to change as less as possible the surface charges of the POM
 so that differences emerge only as a function of the total structural charge. The structure of the POM is kept as described in  Table \ref{POM-SIM}.
 For the systems POM5, POM5-salt, POM0 and POM0-salt, an alternative method to change the total charge is also tested: 
The charge of the phosphorus and the charges of the tungsten were increased (or reduced) by the same amount. 
The radial distribution functions between the POM and water atoms for these two different charge distributions show no significant difference. 
 It should be stressed that POM3 and POM4 ions exist, respectively with the structures [PW$_{12}$O$_{40}$]$^{3-}$ and [SiW$_{12}$O$_{40}$]$^{4-}$. Similar systems with a radius of about $0.5$ nm exist also for charges $-5$~$e$, $-6$~$e$ and higher charges \cite{Cronin2006}. 

The SPC/E water model was chosen with a time step of $\SI{1}{\femto\second}$ in combination with the SHAKE algorithm. 
Several equilibration simulations were run, $\SI{100}{\pico\second}$ in $NVT$, $\SI{1000}{\pico\second}$ in $NpT$ and 
another $\SI{100}{\pico\second}$ in $NVT$. 
Subsequently the POM was fixed in the center of the simulation box setting all forces 
 and its rotational velocity to zero at every time step. 
Another equilibrium run was done to generate
different starting configurations for the non-equilibrium simulations. 
 For all systems, a flow was applied during a $2\,\si{\nano\second}$ simulation, prior to the production run, in order 
to establish steady state conditions. 
These systems were then simulated at least over $37\,\si{\nano\second}$ in total, if we add the simulation time starting from different configurations.
To derive the friction coefficient, we have computed the averaged solvent velocity far from the POM, as described before.
 
\begin{figure}
\includegraphics[scale=0.3]{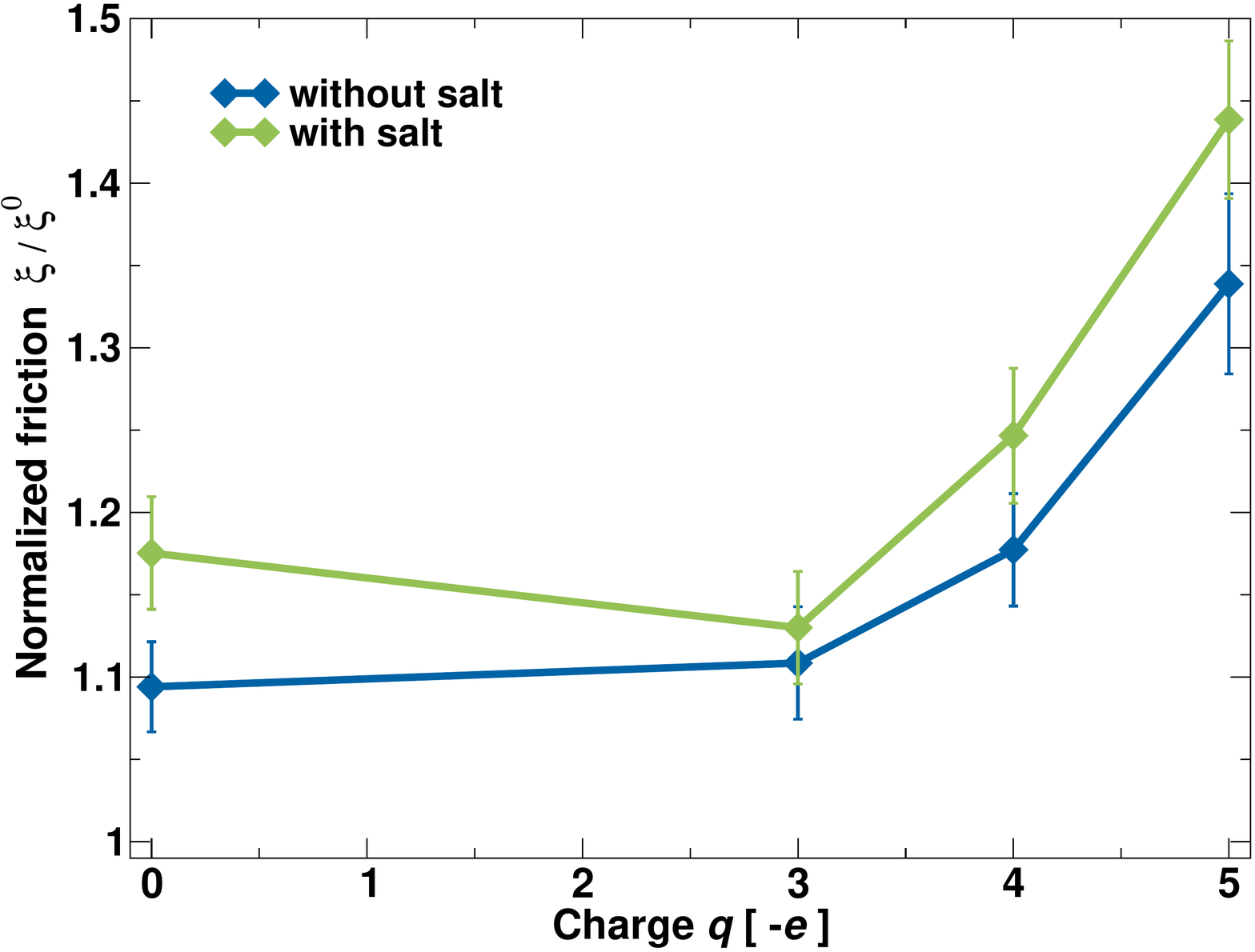}
\includegraphics[scale=0.3]{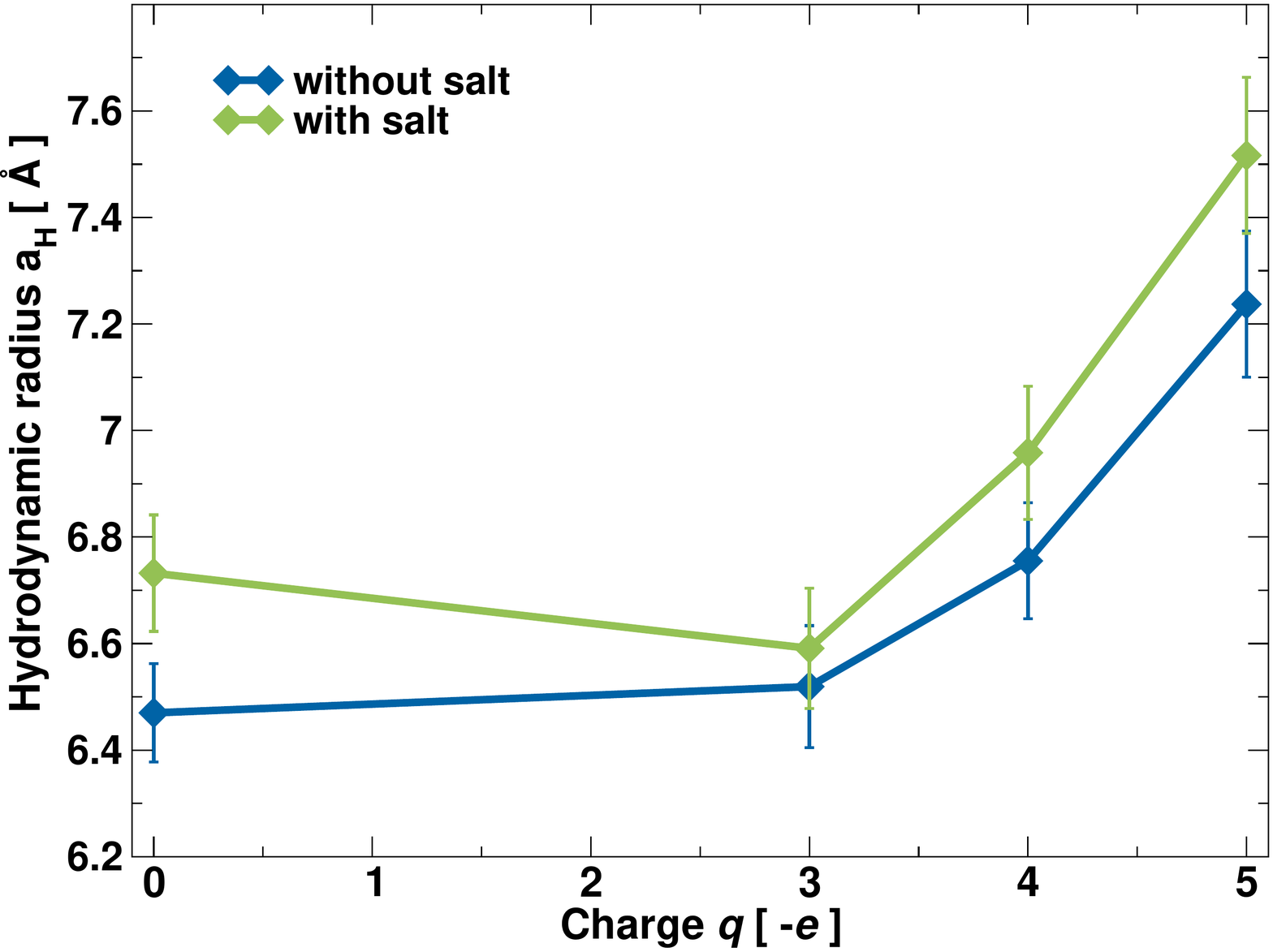}
 \caption{{Friction coefficient (top) and hydrodynamic radius (bottom) of the central nanoparticle as a function of its charge, with and without added salt, computed from non-equilibrium simulations.The error bars correspond to the standard uncertainty.}} 
 \label{friction-compare}

 \end{figure}

The values of the friction coefficient obtained for the different nanoparticles 
are shown in     Fig. \ref{friction-compare} and compared to the values obtained in the presence of an added salt. 
The friction coefficient is divided by $\xi^0$, which is the friction coefficient that would be obtained for an uncharged spherical obstacle of radius equal to $6.0$~\AA~at infinite dilution in water with stick boundary conditions. 
The values of the hydrodynamic radius deduced from the friction coefficient using Hasimoto's result using the method detailed in the methodological section are given in     Fig. \ref{friction-compare}.

The hydrodynamic radius of the neutral nanoparticle is found equal to  $6.45\pm0.01$~\AA. The standard uncertainty is evaluated from the standard uncertainty of the velocity deduced from a block analysis.
 For the POM with charge $-4$ $e$, the hydrodynamic radius is found equal to $6.75\pm0.01$~\AA.  This value is close to our recent experimental study of silicotungstate ions. The electroacoustic signal of aqueous solutions of silicotungstate was indeed measured and analyzed with a novel theoretical treatment in terms of charge and hydrodynamic radius \cite{Pusset2015}: The best fit of the experimental data  corresponds to $6.30$~\AA. 

As it is shown in    Fig. \ref{friction-compare}, the friction coefficient of the nanoparticle increases with increasing charge, and as a consequence the hydrodynamic radius increases with increasing charge. The increase of the computed hydrodynamic radius is significant, from $6.45$~\AA~  for the neutral nanoparticle to $7.24$~\AA~for a charge equal to $-5e$ without salt. It is rather surprising if one considers that the spatial coordinates of the atoms of the nanoparticle do not change at all.

\begin{figure}
\includegraphics[scale=0.4]{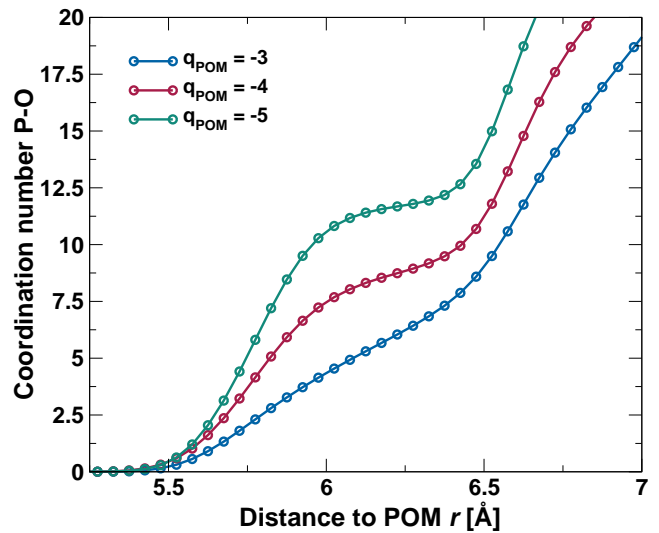}
 \caption{{Number of water molecules as function of the distance to the center of the nanoparticle in the systems with no added salt, computed from the radial distribution function between the phosphorus central atom of the nanoparticle and the oxygen atom of the water molecule.}}
 \label{NdeR}
\end{figure}

\begin{figure}
\includegraphics[scale=0.4]{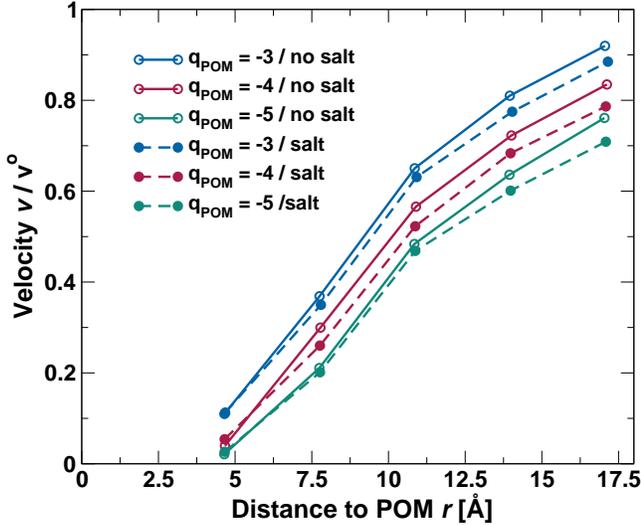}
 \caption{{Mean velocity of water molecules as function of the distance to the center of the nanoparticle.}}
 \label{modulus}
 \end{figure}

For such small nanoparticles, it is difficult to establish exactly what should be the value of the hydrodynamic radius from the microscopic structure. 
Nevertheless, it should be related to the average distance between the water molecules in the first solvation shell containing mobile water molecules and the center of the nanoparticle.
Two effects can explain the variations of the hydrodynamic radius. First, water molecules are more attracted by the nanoparticle when its charge increases. 
This can be quantitatively observed in    Fig. \ref{NdeR} which gives the number of water molecules as a function of the distance to the phosphorus atom of the POM. 
The number of water molecules as a function of the distance has been computed as an integral of the radial distribution function. 
The plateau in the coordination number plot as function of the distance for the systems POM4, POM5 indicates that the POM is solvated respectively by 8, and 10 water molecules when its charge increases from 4 to 5. On the contrary, no clear plateau appears for the POM3  which means that the solvation shell is less clearly defined than with larger charges. 
 Therefore, the mobility of water molecules decreases with the charge of the nanoparticle, as it is shown in    Fig. \ref{modulus} that gives the mean velocity of water molecules as a function of the distance to the center of the POM nanoparticle. It appears clearly in    Fig. \ref{modulus} that the velocity of solvent molecules is almost zero ($0.5$ m.s$^{-1}$) for systems POM4 and POM5 at a distance equal to 5 \AA, with or without added salt, whereas it is about three times larger for POM3. 

\begin{figure}
\includegraphics[scale=0.4]{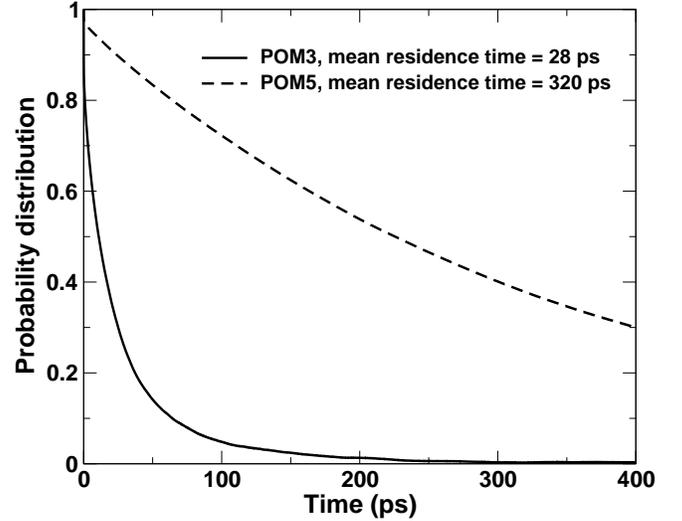}
 \caption{{Residence time of water molecules within a spherical volume of radius  around the nanoparticle.}}
 \label{residence}
 \end{figure}

In order to compare the time spent by a water molecule around a nanoparticle depending on its charge, residence times around 
POM3 and POM5 particles have been evaluated.
The survival probability for a molecule to be found within a given distance $d$ of the center of the nanoparticle 
can be calculated:
$$
P(t)=\frac{\langle S_I(t) \rangle}{\langle S_I(0)\rangle}
$$
where $S_I(t)=1$ if the molecule is present in the sphere of radius $d$ centered around the POM both at time 0 and $t$, 
and $S_I(t)=0$ otherwise. 
Such a definition allows for molecules to leave and come back into the sphere between $0$ and $t$ and characterizes an intermittent survival probability. 
$d$ corresponds to the position of the first minimum of the radial distribution functions between the phosphorus atom and the oxygen atom of water molecules, i.e. $6.2$ \AA.
$P(t)$ can be fitted by a decreasing exponential function $\exp(-t/\tau)$, 
where $\tau$ is the residence time in the volume occupied by the sphere.
More realistic ways to calculate residence times exist, which can be better compare with experiments \cite{Laage}. 
However, in our case, the aim is to see how much the charge of the nanoparticle can slow down water molecules surrounding it. 
Therefore, the comparison between intermittent residence times calculated in the same manner 
already gives interesting insight.  
The probability distribution of residence times in a spherical volume of radius $6.2$ \AA \ 
are shown in    Fig. \ref{residence}. 
The average residence time of water around the nanoparticle is $28$ ns in the case of the nanoparticle  of charge $-3e$ and $320$ ns in the case of the nanoparticle of charge $-5e$. This confirms the structural behavior deduced from the coordination numbers: Water molecules stay a significantly much longer time close to the nanoparticle when it is more charged. As a consequence, it looks like the hydrodynamic radius is larger, accounting for water molecules that would be stuck at the surface of the nanoparticle. In recent simulations of electroosmosis in a slit geometry, such dependence of the flow at the surface of the solid on its charge has also been observed \cite{BarratJCP2014}. 

 A second phenomenon which increases the hydrodynamic radius when the charge increases is the so-called electrostatic friction, 
arising from the interactions with both water molecules (dielectric friction) and small ions. 
The friction is influenced by the long-range electrostatic interactions, which can effectively increase the hydrodynamic radius. The Hubbard-Onsager theory \cite{Hubbard,Wolynes80} yield an estimation of the increase of the friction due to the interaction with water dipoles. In a subsequent study, a detailed comparison of our results with this theory and a more exhaustive analysis of water relaxation around the POM will enable us to unravel whether the increase of the friction with charge can be quantitatively attributed to dielectric friction, or to the strong water adsorption. Please note that Hasimoto's treatment of the flow in an array of spherical obstacles does not take into account the influence of electrostatics on the local viscosity of water. We shall address this limitation in our forthcoming study.    
 
The influence of salt on the friction coefficient is small but systematic: The friction in the presence of added salt is always a few percents larger than the friction without salt. 
As it can be observed in Fig. \ref{modulus}, the presence of added salt leads to a decrease of the velocity of water molecules  compared to the case without salt. The decrease is of about 10\%. This reflects a small  enhancement of the effective viscosity of the solvent due to the presence of the added salt. This effect then slightly impacts the friction coefficient computed from the simulation (the friction increases slightly), which  increases the effective hydrodynamic radius. This indirect effect of salt on the hydrodynamic radius is weak, but it is systematic. In a previous work,  Chowdhuri and Chandra \cite{chowdhuriJCP2001} performed atomistic molecular dynamics simulations of aqueous solutions of potassium chloride with  the same interaction potential as in the present work.   They showed that the dynamics of water molecules slows down as the salt concentration increases. At a KCl concentration of 0.88 mol/L, they observe a decrease of the self-diffusion coefficient of water molecules of about 6\%. We can thus argue that the presence of potassium and chloride ions (at a concentration of 0.5 mol/L in our simulations) impacts the dynamics of water molecules and yields a small increase of the effective viscosity of the solvent.

\section{Conclusion}
\label{conclusion}
Our goal was to find a robust procedure to evaluate the hydrodynamic radius of a nanoparticle in water from numerical simulation. 
Indeed, this quantity is not always easy to deduce from experiments but must be known to perform mesoscopic numerical simulations of the dynamic properties of nanoparticle dispersions. 
Moreover, such study is very important to determine whether a specific nanoparticle is sufficiently large to behave as a typical colloidal particle from the dynamical point of view. For instance, the existence of a linear dependence between the size of the particle and its friction, or its diffusion coefficient, can be evaluated, 
{\it i. e.} the relevance of the Stokes-Einstein relationship can be evaluated \cite{ZhangJPCB2016}. 
Small nanoparticles are indeed in between colloids and electrolytes, and for electrolytes, the link between the structure (size, shape) and the friction is much less clear than for colloidal particles.
      
We have proposed to use non-equilibrium molecular dynamics simulations to study the flow of water around the nanoparticle, and then to compute the friction coefficient of the nanoparticle. 
One difficulty in non-equilibrium simulations is to find a good way to induce the flow without heating the system. 
As we have shown in the present study, this difficulty is more pronounced for the flow around a spherical nanoparticle because the geometry of the velocity field is less symmetric than in a channel.
We compared two procedures to thermostat the simulation box, both available as options in common open-source molecular dynamics simulation packages. 
The first one, the PUT, divides the box volume into small bins, in which the average velocity of the solvent molecules is approximately uniform.
 The second one consists in thermostatting the system only in a subpart of the simulation box, where the velocity field is almost uniform. We have shown that for the flow around a roughly spherical object the choice of a cylindrical shape 
 of the boundary between the thermostatted volume and the rest of the box allows one to obtain
 a constant temperature in all the simulation box. 

We have then performed non-equilibrium simulations of the flow of water around nanoparticles of different charges using a partial thermostat of cylindrical shape. The atomic structure of the nanoparticle represents the one of typical polyoxometalate ions (POM). The friction coefficient of the nanoparticle was deduced from the average solvent velocity with a good precision.
This result enabled us to compute the hydrodynamic radius by using Hasimoto's approximate analytical result for the flow across a cubic array of spheres. 
We found that the hydrodynamic radius significantly increases with the charge of the nanoparticle, a phenomenon that had not been quantified so far using molecular dynamics. This result suggests that such nanoparticle should be seen as a big ion more than as a small colloidal particle.  
It should 
be of particular interest for the analysis of routine experiments on nanoparticles, such as dynamic light scattering or laser zetametry when the diffusion coefficient or the electrophoretic mobility are generally evaluated 
using the Stokes-Einstein estimation of friction, which only depends on the radius of the nanoparticle and not directly on its charge.  
While more sophisticated theories can be used, such as the Hubbard-Onsager theory of dielectric friction,  their validity as a quantitative tool to analyse experimental data is questionable. Molecular dynamics provides a much more precise and convincing way to quantify the deviations from the Stokes-Einstein law. 
A forthcoming study should enable us to relate these quantitative results with existing analytical theories, and to get a clearer view of the relevant physical phenomena to take into account to predict the effective hydrodynamic radius or diffusion coefficient of a small charged nanoparticle.   

\vspace{1cm}
\noindent {\bf Acknowledgement}
The work has received funding from the European Union's Horizon 2020 research and innovation programme under grant agreement No 674979-NANOTRANS. 

\begin{thebibliography}{50}%
\makeatletter
\providecommand \@ifxundefined [1]{%
 \@ifx{#1\undefined}
}%
\providecommand \@ifnum [1]{%
 \ifnum #1\expandafter \@firstoftwo
 \else \expandafter \@secondoftwo
 \fi
}%
\providecommand \@ifx [1]{%
 \ifx #1\expandafter \@firstoftwo
 \else \expandafter \@secondoftwo
 \fi
}%
\providecommand \natexlab [1]{#1}%
\providecommand \enquote  [1]{``#1''}%
\providecommand \bibnamefont  [1]{#1}%
\providecommand \bibfnamefont [1]{#1}%
\providecommand \citenamefont [1]{#1}%
\providecommand \href@noop [0]{\@secondoftwo}%
\providecommand \href [0]{\begingroup \@sanitize@url \@href}%
\providecommand \@href[1]{\@@startlink{#1}\@@href}%
\providecommand \@@href[1]{\endgroup#1\@@endlink}%
\providecommand \@sanitize@url [0]{\catcode `\\12\catcode `\$12\catcode
  `\&12\catcode `\#12\catcode `\^12\catcode `\_12\catcode `\%12\relax}%
\providecommand \@@startlink[1]{}%
\providecommand \@@endlink[0]{}%
\providecommand \url  [0]{\begingroup\@sanitize@url \@url }%
\providecommand \@url [1]{\endgroup\@href {#1}{\urlprefix }}%
\providecommand \urlprefix  [0]{URL }%
\providecommand \Eprint [0]{\href }%
\providecommand \doibase [0]{http://dx.doi.org/}%
\providecommand \selectlanguage [0]{\@gobble}%
\providecommand \bibinfo  [0]{\@secondoftwo}%
\providecommand \bibfield  [0]{\@secondoftwo}%
\providecommand \translation [1]{[#1]}%
\providecommand \BibitemOpen [0]{}%
\providecommand \bibitemStop [0]{}%
\providecommand \bibitemNoStop [0]{.\EOS\space}%
\providecommand \EOS [0]{\spacefactor3000\relax}%
\providecommand \BibitemShut  [1]{\csname bibitem#1\endcsname}%
\let\auto@bib@innerbib\@empty
\bibitem [{\citenamefont {Stark}\ \emph {et~al.}(2015)\citenamefont {Stark},
  \citenamefont {Stoessel}, \citenamefont {Wohlleben},\ and\ \citenamefont
  {Hafner}}]{Stark2015}%
  \BibitemOpen
  \bibfield  {author} {\bibinfo {author} {\bibfnamefont {W.~J.}\ \bibnamefont
  {Stark}}, \bibinfo {author} {\bibfnamefont {P.~R.}\ \bibnamefont {Stoessel}},
  \bibinfo {author} {\bibfnamefont {W.}~\bibnamefont {Wohlleben}}, \ and\
  \bibinfo {author} {\bibfnamefont {A.}~\bibnamefont {Hafner}},\ }\href
  {\doibase 10.1039/C4CS00362D} {\bibfield  {journal} {\bibinfo  {journal}
  {Chem. Soc. Rev.}\ }\textbf {\bibinfo {volume} {44}},\ \bibinfo {pages}
  {5793} (\bibinfo {year} {2015})}\BibitemShut {NoStop}%
\bibitem [{\citenamefont {Ermak}\ and\ \citenamefont
  {McCammon}(1978)}]{ErmakJCP78}%
  \BibitemOpen
  \bibfield  {author} {\bibinfo {author} {\bibfnamefont {D.~L.}\ \bibnamefont
  {Ermak}}\ and\ \bibinfo {author} {\bibfnamefont {J.~A.}\ \bibnamefont
  {McCammon}},\ }\href@noop {} {\bibfield  {journal} {\bibinfo  {journal} {J.
  Chem. Phys.}\ }\textbf {\bibinfo {volume} {69}},\ \bibinfo {pages} {1352}
  (\bibinfo {year} {1978})}\BibitemShut {NoStop}%
\bibitem [{\citenamefont {Dahirel}\ \emph {et~al.}(2009)\citenamefont
  {Dahirel}, \citenamefont {Jardat}, \citenamefont {Dufr\^eche},\ and\
  \citenamefont {Turq}}]{dahirel_JCP_2009}%
  \BibitemOpen
  \bibfield  {author} {\bibinfo {author} {\bibfnamefont {V.}~\bibnamefont
  {Dahirel}}, \bibinfo {author} {\bibfnamefont {M.}~\bibnamefont {Jardat}},
  \bibinfo {author} {\bibfnamefont {J.~F.}\ \bibnamefont {Dufr\^eche}}, \ and\
  \bibinfo {author} {\bibfnamefont {P.}~\bibnamefont {Turq}},\ }\href@noop {}
  {\bibfield  {journal} {\bibinfo  {journal} {J. Chem. Phys.}\ }\textbf
  {\bibinfo {volume} {131}},\ \bibinfo {pages} {234105} (\bibinfo {year}
  {2009})}\BibitemShut {NoStop}%
\bibitem [{\citenamefont {Padding}\ and\ \citenamefont
  {Louis}(2006)}]{Padding2006}%
  \BibitemOpen
  \bibfield  {author} {\bibinfo {author} {\bibfnamefont {J.~T.}\ \bibnamefont
  {Padding}}\ and\ \bibinfo {author} {\bibfnamefont {A.~A.}\ \bibnamefont
  {Louis}},\ }\href {\doibase 10.1103/PhysRevE.74.031402} {\bibfield  {journal}
  {\bibinfo  {journal} {Phys. Rev. E}\ }\textbf {\bibinfo {volume} {74}},\
  \bibinfo {pages} {1} (\bibinfo {year} {2006})},\ \Eprint
  {http://arxiv.org/abs/0603391} {arXiv:0603391 [cond-mat]} \BibitemShut
  {NoStop}%
\bibitem [{\citenamefont {Gompper}\ \emph {et~al.}(2008)\citenamefont
  {Gompper}, \citenamefont {Ihle}, \citenamefont {Kroll},\ and\ \citenamefont
  {Winkler}}]{gompper_multi-particle_2008}%
  \BibitemOpen
  \bibfield  {author} {\bibinfo {author} {\bibfnamefont {G.}~\bibnamefont
  {Gompper}}, \bibinfo {author} {\bibfnamefont {T.}~\bibnamefont {Ihle}},
  \bibinfo {author} {\bibfnamefont {D.}~\bibnamefont {Kroll}}, \ and\ \bibinfo
  {author} {\bibfnamefont {R.}~\bibnamefont {Winkler}},\ }\href@noop {}
  {\bibfield  {journal} {\bibinfo  {journal} {Advances in Polymer Science}\
  }\textbf {\bibinfo {volume} {221}},\ \bibinfo {pages} {1} (\bibinfo {year}
  {2008})}\BibitemShut {NoStop}%
\bibitem [{\citenamefont {Espa\~{n}ol}\ and\ \citenamefont
  {Warren}(2017)}]{espanol2017}%
  \BibitemOpen
  \bibfield  {author} {\bibinfo {author} {\bibfnamefont {P.}~\bibnamefont
  {Espa\~{n}ol}}\ and\ \bibinfo {author} {\bibfnamefont {P.~B.}\ \bibnamefont
  {Warren}},\ }\href {\doibase 10.1063/1.4979514} {\bibfield  {journal}
  {\bibinfo  {journal} {J. Chem. Phys.}\ }\textbf {\bibinfo {volume} {146}},\
  \bibinfo {pages} {150901} (\bibinfo {year} {2017})},\ \Eprint
  {http://arxiv.org/abs/http://dx.doi.org/10.1063/1.4979514}
  {http://dx.doi.org/10.1063/1.4979514} \BibitemShut {NoStop}%
\bibitem [{\citenamefont {Kops-Werkhoven}\ \emph {et~al.}(1982)\citenamefont
  {Kops-Werkhoven}, \citenamefont {Pathmamanoharan}, \citenamefont {Vrij},\
  and\ \citenamefont {Fijnaut}}]{Kops1982}%
  \BibitemOpen
  \bibfield  {author} {\bibinfo {author} {\bibfnamefont {M.~M.}\ \bibnamefont
  {Kops-Werkhoven}}, \bibinfo {author} {\bibfnamefont {C.}~\bibnamefont
  {Pathmamanoharan}}, \bibinfo {author} {\bibfnamefont {A.}~\bibnamefont
  {Vrij}}, \ and\ \bibinfo {author} {\bibfnamefont {H.~M.}\ \bibnamefont
  {Fijnaut}},\ }\href {\doibase 10.1063/1.443864} {\bibfield  {journal}
  {\bibinfo  {journal} {J. Chem. Phys.}\ }\textbf {\bibinfo {volume} {77}},\
  \bibinfo {pages} {5913} (\bibinfo {year} {1982})}\BibitemShut {NoStop}%
\bibitem [{\citenamefont {Schmidt}\ and\ \citenamefont
  {Skinner}(2003)}]{SchmidtJCP2003}%
  \BibitemOpen
  \bibfield  {author} {\bibinfo {author} {\bibfnamefont {J.~R.}\ \bibnamefont
  {Schmidt}}\ and\ \bibinfo {author} {\bibfnamefont {J.~L.}\ \bibnamefont
  {Skinner}},\ }\href@noop {} {\bibfield  {journal} {\bibinfo  {journal} {J.
  Chem. Phys.}\ }\textbf {\bibinfo {volume} {119}},\ \bibinfo {pages} {8062}
  (\bibinfo {year} {2003})}\BibitemShut {NoStop}%
\bibitem [{\citenamefont {Schmidt}\ and\ \citenamefont
  {Skinner}(2004)}]{SchmidtJPCB2004}%
  \BibitemOpen
  \bibfield  {author} {\bibinfo {author} {\bibfnamefont {J.~R.}\ \bibnamefont
  {Schmidt}}\ and\ \bibinfo {author} {\bibfnamefont {J.~L.}\ \bibnamefont
  {Skinner}},\ }\href@noop {} {\bibfield  {journal} {\bibinfo  {journal} {J.
  Phys. Chem. B}\ }\textbf {\bibinfo {volume} {108}},\ \bibinfo {pages} {6767}
  (\bibinfo {year} {2004})}\BibitemShut {NoStop}%
\bibitem [{\citenamefont {Morrone}\ \emph {et~al.}(2012)\citenamefont
  {Morrone}, \citenamefont {Li},\ and\ \citenamefont {Berne}}]{BerneJPCB2012}%
  \BibitemOpen
  \bibfield  {author} {\bibinfo {author} {\bibfnamefont {J.}~\bibnamefont
  {Morrone}}, \bibinfo {author} {\bibfnamefont {J.}~\bibnamefont {Li}}, \ and\
  \bibinfo {author} {\bibfnamefont {B.~J.}\ \bibnamefont {Berne}},\ }\href@noop
  {} {\bibfield  {journal} {\bibinfo  {journal} {J. Phys. Chem. B}\ }\textbf
  {\bibinfo {volume} {116}},\ \bibinfo {pages} {378} (\bibinfo {year}
  {2012})}\BibitemShut {NoStop}%
\bibitem [{\citenamefont {Ishii}\ and\ \citenamefont
  {Ohtori}(2016)}]{IshiiPRE2016}%
  \BibitemOpen
  \bibfield  {author} {\bibinfo {author} {\bibfnamefont {Y.}~\bibnamefont
  {Ishii}}\ and\ \bibinfo {author} {\bibfnamefont {N.}~\bibnamefont {Ohtori}},\
  }\href@noop {} {\bibfield  {journal} {\bibinfo  {journal} {Phys. Rev. E}\
  }\textbf {\bibinfo {volume} {93}},\ \bibinfo {pages} {050104(R)} (\bibinfo
  {year} {2016})}\BibitemShut {NoStop}%
\bibitem [{\citenamefont {Pau}\ \emph {et~al.}(1990)\citenamefont {Pau},
  \citenamefont {Berg},\ and\ \citenamefont {McMillan}}]{pauJPC90}%
  \BibitemOpen
  \bibfield  {author} {\bibinfo {author} {\bibfnamefont {P.~C.~F.}\
  \bibnamefont {Pau}}, \bibinfo {author} {\bibfnamefont {J.~O.}\ \bibnamefont
  {Berg}}, \ and\ \bibinfo {author} {\bibfnamefont {W.~G.}\ \bibnamefont
  {McMillan}},\ }\href@noop {} {\bibfield  {journal} {\bibinfo  {journal} {J.
  Phys. Chem.}\ }\textbf {\bibinfo {volume} {94}},\ \bibinfo {pages} {2671}
  (\bibinfo {year} {1990})}\BibitemShut {NoStop}%
\bibitem [{\citenamefont {Hubbard}\ and\ \citenamefont
  {Onsager}(1977)}]{Hubbard}%
  \BibitemOpen
  \bibfield  {author} {\bibinfo {author} {\bibfnamefont {J.}~\bibnamefont
  {Hubbard}}\ and\ \bibinfo {author} {\bibfnamefont {L.}~\bibnamefont
  {Onsager}},\ }\href@noop {} {\bibfield  {journal} {\bibinfo  {journal} {J.
  Chem. Phys.}\ }\textbf {\bibinfo {volume} {67}},\ \bibinfo {pages} {4850}
  (\bibinfo {year} {1977})}\BibitemShut {NoStop}%
\bibitem [{\citenamefont {Wolynes}(1980)}]{Wolynes80}%
  \BibitemOpen
  \bibfield  {author} {\bibinfo {author} {\bibfnamefont {P.~G.}\ \bibnamefont
  {Wolynes}},\ }\href@noop {} {\bibfield  {journal} {\bibinfo  {journal} {Ann.
  Rev. Phys. Chem.}\ }\textbf {\bibinfo {volume} {31}},\ \bibinfo {pages} {345}
  (\bibinfo {year} {1980})}\BibitemShut {NoStop}%
\bibitem [{\citenamefont {Yoshida}\ \emph {et~al.}(2014)\citenamefont
  {Yoshida}, \citenamefont {Mizuno}, \citenamefont {Kinjo}, \citenamefont
  {Washizu},\ and\ \citenamefont {Barrat}}]{BarratJCP2014}%
  \BibitemOpen
  \bibfield  {author} {\bibinfo {author} {\bibfnamefont {H.}~\bibnamefont
  {Yoshida}}, \bibinfo {author} {\bibfnamefont {H.}~\bibnamefont {Mizuno}},
  \bibinfo {author} {\bibfnamefont {T.}~\bibnamefont {Kinjo}}, \bibinfo
  {author} {\bibfnamefont {H.}~\bibnamefont {Washizu}}, \ and\ \bibinfo
  {author} {\bibfnamefont {J.-L.}\ \bibnamefont {Barrat}},\ }\href@noop {}
  {\bibfield  {journal} {\bibinfo  {journal} {J. Chem. Phys.}\ }\textbf
  {\bibinfo {volume} {140}},\ \bibinfo {pages} {214701} (\bibinfo {year}
  {2014})}\BibitemShut {NoStop}%
\bibitem [{\citenamefont {Bhadauria}\ and\ \citenamefont
  {Aluru}(2017)}]{AluruJCP2017A}%
  \BibitemOpen
  \bibfield  {author} {\bibinfo {author} {\bibfnamefont {R.}~\bibnamefont
  {Bhadauria}}\ and\ \bibinfo {author} {\bibfnamefont {N.~R.}\ \bibnamefont
  {Aluru}},\ }\href@noop {} {\bibfield  {journal} {\bibinfo  {journal} {J.
  Chem. Phys.}\ }\textbf {\bibinfo {volume} {146}},\ \bibinfo {pages} {184106}
  (\bibinfo {year} {2017})}\BibitemShut {NoStop}%
\bibitem [{\citenamefont {Joly}\ \emph {et~al.}(2006)\citenamefont {Joly},
  \citenamefont {Ybert}, \citenamefont {Trizac},\ and\ \citenamefont
  {Bocquet}}]{JolyJCP2006}%
  \BibitemOpen
  \bibfield  {author} {\bibinfo {author} {\bibfnamefont {L.}~\bibnamefont
  {Joly}}, \bibinfo {author} {\bibfnamefont {C.}~\bibnamefont {Ybert}},
  \bibinfo {author} {\bibfnamefont {E.}~\bibnamefont {Trizac}}, \ and\ \bibinfo
  {author} {\bibfnamefont {L.}~\bibnamefont {Bocquet}},\ }\href {\doibase
  10.1063/1.2397677} {\bibfield  {journal} {\bibinfo  {journal} {J. Chem.
  Phys.}\ }\textbf {\bibinfo {volume} {125}},\ \bibinfo {pages} {204716}
  (\bibinfo {year} {2006})},\ \Eprint
  {http://arxiv.org/abs/http://dx.doi.org/10.1063/1.2397677}
  {http://dx.doi.org/10.1063/1.2397677} \BibitemShut {NoStop}%
\bibitem [{\citenamefont {Botan}\ \emph {et~al.}(2011)\citenamefont {Botan},
  \citenamefont {Rotenberg}, \citenamefont {Marry}, \citenamefont {Turq},\ and\
  \citenamefont {Noetinger}}]{Botan2011}%
  \BibitemOpen
  \bibfield  {author} {\bibinfo {author} {\bibfnamefont {A.}~\bibnamefont
  {Botan}}, \bibinfo {author} {\bibfnamefont {B.}~\bibnamefont {Rotenberg}},
  \bibinfo {author} {\bibfnamefont {V.}~\bibnamefont {Marry}}, \bibinfo
  {author} {\bibfnamefont {P.}~\bibnamefont {Turq}}, \ and\ \bibinfo {author}
  {\bibfnamefont {B.}~\bibnamefont {Noetinger}},\ }\href {\doibase
  10.1021/jp204772c} {\bibfield  {journal} {\bibinfo  {journal} {J. Phys. Chem.
  C}\ }\textbf {\bibinfo {volume} {115}},\ \bibinfo {pages} {16109} (\bibinfo
  {year} {2011})}\BibitemShut {NoStop}%
\bibitem [{\citenamefont {Botan}\ \emph {et~al.}(2013)\citenamefont {Botan},
  \citenamefont {Marry}, \citenamefont {Rotenberg}, \citenamefont {Turq},\ and\
  \citenamefont {Noetinger}}]{botan_how_2013}%
  \BibitemOpen
  \bibfield  {author} {\bibinfo {author} {\bibfnamefont {A.}~\bibnamefont
  {Botan}}, \bibinfo {author} {\bibfnamefont {V.}~\bibnamefont {Marry}},
  \bibinfo {author} {\bibfnamefont {B.}~\bibnamefont {Rotenberg}}, \bibinfo
  {author} {\bibfnamefont {P.}~\bibnamefont {Turq}}, \ and\ \bibinfo {author}
  {\bibfnamefont {B.}~\bibnamefont {Noetinger}},\ }\href@noop {} {\bibfield
  {journal} {\bibinfo  {journal} {J. Phys. Chem. C}\ }\textbf {\bibinfo
  {volume} {117}},\ \bibinfo {pages} {978} (\bibinfo {year}
  {2013})}\BibitemShut {NoStop}%
\bibitem [{\citenamefont {Bonthuis}\ and\ \citenamefont
  {Netz}(2013)}]{Bonthuis2013}%
  \BibitemOpen
  \bibfield  {author} {\bibinfo {author} {\bibfnamefont {D.~J.}\ \bibnamefont
  {Bonthuis}}\ and\ \bibinfo {author} {\bibfnamefont {R.~R.}\ \bibnamefont
  {Netz}},\ }\href {\doibase 10.1021/jp402482q} {\bibfield  {journal} {\bibinfo
   {journal} {J. Phys. Chem. B}\ }\textbf {\bibinfo {volume} {117}},\ \bibinfo
  {pages} {11397} (\bibinfo {year} {2013})},\ \Eprint
  {http://arxiv.org/abs/http://dx.doi.org/10.1021/jp402482q}
  {http://dx.doi.org/10.1021/jp402482q} \BibitemShut {NoStop}%
\bibitem [{\citenamefont {Bernardi}\ \emph {et~al.}(2010)\citenamefont
  {Bernardi}, \citenamefont {Todd}, \citenamefont {Hansen}, \citenamefont
  {Searles},\ and\ \citenamefont {Frascoli}}]{Bernardi2010}%
  \BibitemOpen
  \bibfield  {author} {\bibinfo {author} {\bibfnamefont {S.}~\bibnamefont
  {Bernardi}}, \bibinfo {author} {\bibfnamefont {B.~D.}\ \bibnamefont {Todd}},
  \bibinfo {author} {\bibfnamefont {J.~S.}\ \bibnamefont {Hansen}}, \bibinfo
  {author} {\bibfnamefont {D.}~\bibnamefont {Searles}}, \ and\ \bibinfo
  {author} {\bibfnamefont {F.}~\bibnamefont {Frascoli}},\ }\href@noop {}
  {\bibfield  {journal} {\bibinfo  {journal} {J. Chem. Phys.}\ }\textbf
  {\bibinfo {volume} {132}},\ \bibinfo {pages} {244508} (\bibinfo {year}
  {2010})}\BibitemShut {NoStop}%
\bibitem [{\citenamefont {Yong}\ and\ \citenamefont {Zhang}(2013)}]{Yong2013}%
  \BibitemOpen
  \bibfield  {author} {\bibinfo {author} {\bibfnamefont {X.}~\bibnamefont
  {Yong}}\ and\ \bibinfo {author} {\bibfnamefont {L.~T.}\ \bibnamefont
  {Zhang}},\ }\href@noop {} {\bibfield  {journal} {\bibinfo  {journal} {J.
  Chem. Phys.}\ }\textbf {\bibinfo {volume} {138}},\ \bibinfo {pages} {084503}
  (\bibinfo {year} {2013})}\BibitemShut {NoStop}%
\bibitem [{\citenamefont {De~Luca}\ \emph {et~al.}(2014)\citenamefont
  {De~Luca}, \citenamefont {Todd}, \citenamefont {Hansen},\ and\ \citenamefont
  {Daivis}}]{DeLuca2014}%
  \BibitemOpen
  \bibfield  {author} {\bibinfo {author} {\bibfnamefont {S.}~\bibnamefont
  {De~Luca}}, \bibinfo {author} {\bibfnamefont {B.~D.}\ \bibnamefont {Todd}},
  \bibinfo {author} {\bibfnamefont {J.~S.}\ \bibnamefont {Hansen}}, \ and\
  \bibinfo {author} {\bibfnamefont {P.~J.}\ \bibnamefont {Daivis}},\
  }\href@noop {} {\bibfield  {journal} {\bibinfo  {journal} {J. Chem. Phys.}\
  }\textbf {\bibinfo {volume} {140}},\ \bibinfo {pages} {054502} (\bibinfo
  {year} {2014})}\BibitemShut {NoStop}%
\bibitem [{\citenamefont {Evans}\ and\ \citenamefont
  {Morriss}(1984)}]{Evans1984a}%
  \BibitemOpen
  \bibfield  {author} {\bibinfo {author} {\bibfnamefont {J.~D.}\ \bibnamefont
  {Evans}}\ and\ \bibinfo {author} {\bibfnamefont {G.}~\bibnamefont
  {Morriss}},\ }\href@noop {} {\bibfield  {journal} {\bibinfo  {journal} {Phys.
  Rev. A}\ }\textbf {\bibinfo {volume} {30}},\ \bibinfo {pages} {1528}
  (\bibinfo {year} {1984})}\BibitemShut {NoStop}%
\bibitem [{\citenamefont {Martyna}\ \emph {et~al.}(1992)\citenamefont
  {Martyna}, \citenamefont {Klein},\ and\ \citenamefont
  {Tuckerman}}]{NoseHoover}%
  \BibitemOpen
  \bibfield  {author} {\bibinfo {author} {\bibfnamefont {G.~J.}\ \bibnamefont
  {Martyna}}, \bibinfo {author} {\bibfnamefont {M.~L.}\ \bibnamefont {Klein}},
  \ and\ \bibinfo {author} {\bibfnamefont {M.}~\bibnamefont {Tuckerman}},\
  }\href {\doibase 10.1063/1.463940} {\bibfield  {journal} {\bibinfo  {journal}
  {J. Chem. Phys.}\ }\textbf {\bibinfo {volume} {97}},\ \bibinfo {pages} {2635}
  (\bibinfo {year} {1992})}\BibitemShut {NoStop}%
\bibitem [{\citenamefont {Hasimoto}(1959)}]{Hasimoto}%
  \BibitemOpen
  \bibfield  {author} {\bibinfo {author} {\bibfnamefont {H.}~\bibnamefont
  {Hasimoto}},\ }\href@noop {} {\bibfield  {journal} {\bibinfo  {journal} {J.
  Fluid. Mech.}\ }\textbf {\bibinfo {volume} {5}},\ \bibinfo {pages} {317}
  (\bibinfo {year} {1959})}\BibitemShut {NoStop}%
\bibitem [{\citenamefont {Yeh}\ and\ \citenamefont {Hummer}(2004)}]{Yeh2004}%
  \BibitemOpen
  \bibfield  {author} {\bibinfo {author} {\bibfnamefont {I.~C.}\ \bibnamefont
  {Yeh}}\ and\ \bibinfo {author} {\bibfnamefont {G.}~\bibnamefont {Hummer}},\
  }\href {\doibase 10.1021/jp0477147} {\bibfield  {journal} {\bibinfo
  {journal} {J. Phys. Chem. B}\ }\textbf {\bibinfo {volume} {108}},\ \bibinfo
  {pages} {15873} (\bibinfo {year} {2004})}\BibitemShut {NoStop}%
\bibitem [{\citenamefont {Dzubiella}\ and\ \citenamefont
  {Hansen}(2004)}]{Dzubiella2004}%
  \BibitemOpen
  \bibfield  {author} {\bibinfo {author} {\bibfnamefont {J.}~\bibnamefont
  {Dzubiella}}\ and\ \bibinfo {author} {\bibfnamefont {J.~P.}\ \bibnamefont
  {Hansen}},\ }\href@noop {} {\bibfield  {journal} {\bibinfo  {journal} {J.
  Chem. Phys.}\ }\textbf {\bibinfo {volume} {121}},\ \bibinfo {pages} {5514}
  (\bibinfo {year} {2004})}\BibitemShut {NoStop}%
\bibitem [{\citenamefont {Hodne}\ and\ \citenamefont
  {Beattie}(2001)}]{Hodne2001}%
  \BibitemOpen
  \bibfield  {author} {\bibinfo {author} {\bibfnamefont {H.}~\bibnamefont
  {Hodne}}\ and\ \bibinfo {author} {\bibfnamefont {J.~K.}\ \bibnamefont
  {Beattie}},\ }\href@noop {} {\bibfield  {journal} {\bibinfo  {journal}
  {Langmuir}\ }\textbf {\bibinfo {volume} {17}},\ \bibinfo {pages} {3044}
  (\bibinfo {year} {2001})}\BibitemShut {NoStop}%
\bibitem [{\citenamefont {Maestre}\ \emph {et~al.}(2001)\citenamefont
  {Maestre}, \citenamefont {Lopez}, \citenamefont {Bo}, \citenamefont
  {Poblet},\ and\ \citenamefont {Casa\~{n}-Pastor}}]{Maestre2001}%
  \BibitemOpen
  \bibfield  {author} {\bibinfo {author} {\bibfnamefont {J.~M.}\ \bibnamefont
  {Maestre}}, \bibinfo {author} {\bibfnamefont {X.}~\bibnamefont {Lopez}},
  \bibinfo {author} {\bibfnamefont {C.}~\bibnamefont {Bo}}, \bibinfo {author}
  {\bibfnamefont {J.-M.}\ \bibnamefont {Poblet}}, \ and\ \bibinfo {author}
  {\bibfnamefont {N.}~\bibnamefont {Casa\~{n}-Pastor}},\ }\href@noop {}
  {\bibfield  {journal} {\bibinfo  {journal} {J. Am. Chem. Soc.}\ }\textbf
  {\bibinfo {volume} {123}},\ \bibinfo {pages} {3749} (\bibinfo {year}
  {2001})}\BibitemShut {NoStop}%
\bibitem [{\citenamefont {L\'{o}pez}\ \emph {et~al.}(2005)\citenamefont
  {L\'{o}pez}, \citenamefont {Nieto-Draghi}, \citenamefont {Bo}, \citenamefont
  {Avalos},\ and\ \citenamefont {Poblet}}]{Lopez2005b}%
  \BibitemOpen
  \bibfield  {author} {\bibinfo {author} {\bibfnamefont {X.}~\bibnamefont
  {L\'{o}pez}}, \bibinfo {author} {\bibfnamefont {C.}~\bibnamefont
  {Nieto-Draghi}}, \bibinfo {author} {\bibfnamefont {C.}~\bibnamefont {Bo}},
  \bibinfo {author} {\bibfnamefont {J.~B.}\ \bibnamefont {Avalos}}, \ and\
  \bibinfo {author} {\bibfnamefont {J.~M.}\ \bibnamefont {Poblet}},\ }\href
  {\doibase 10.1021/jp046862u} {\bibfield  {journal} {\bibinfo  {journal} {J.
  Phys. Chem. A}\ }\textbf {\bibinfo {volume} {109}},\ \bibinfo {pages} {1216}
  (\bibinfo {year} {2005})}\BibitemShut {NoStop}%
\bibitem [{\citenamefont {Leroy}\ \emph {et~al.}(2008)\citenamefont {Leroy},
  \citenamefont {Mir\'{o}}, \citenamefont {Poblet}, \citenamefont {Bo},\ and\
  \citenamefont {{Bonet Avalos}}}]{Leroy2008b}%
  \BibitemOpen
  \bibfield  {author} {\bibinfo {author} {\bibfnamefont {F.}~\bibnamefont
  {Leroy}}, \bibinfo {author} {\bibfnamefont {P.}~\bibnamefont {Mir\'{o}}},
  \bibinfo {author} {\bibfnamefont {J.~M.}\ \bibnamefont {Poblet}}, \bibinfo
  {author} {\bibfnamefont {C.}~\bibnamefont {Bo}}, \ and\ \bibinfo {author}
  {\bibfnamefont {J.}~\bibnamefont {{Bonet Avalos}}},\ }\href {\doibase
  10.1021/jp077098p} {\bibfield  {journal} {\bibinfo  {journal} {J. Phys. Chem.
  B}\ }\textbf {\bibinfo {volume} {112}},\ \bibinfo {pages} {8591} (\bibinfo
  {year} {2008})}\BibitemShut {NoStop}%
\bibitem [{\citenamefont {Chaumont}\ and\ \citenamefont
  {Wipff}(2008)}]{Chaumont2008b}%
  \BibitemOpen
  \bibfield  {author} {\bibinfo {author} {\bibfnamefont {A.}~\bibnamefont
  {Chaumont}}\ and\ \bibinfo {author} {\bibfnamefont {G.}~\bibnamefont
  {Wipff}},\ }\href {\doibase 10.1039/b810440a} {\bibfield  {journal} {\bibinfo
   {journal} {Phys. Chem. Chem. Phys.}\ }\textbf {\bibinfo {volume} {10}},\
  \bibinfo {pages} {6940} (\bibinfo {year} {2008})}\BibitemShut {NoStop}%
\bibitem [{\citenamefont {Chaumont}\ and\ \citenamefont
  {Wipff}(2013)}]{Chaumont2013b}%
  \BibitemOpen
  \bibfield  {author} {\bibinfo {author} {\bibfnamefont {A.}~\bibnamefont
  {Chaumont}}\ and\ \bibinfo {author} {\bibfnamefont {G.}~\bibnamefont
  {Wipff}},\ }\href {\doibase 10.1002/ejic.201200883} {\bibfield  {journal}
  {\bibinfo  {journal} {Eur. J. Inorg. Chem}\ ,\ \bibinfo {pages} {1835}}
  (\bibinfo {year} {2013})}\BibitemShut {NoStop}%
\bibitem [{\citenamefont {Noe-Spirlet}\ \emph {et~al.}(1975)\citenamefont
  {Noe-Spirlet}, \citenamefont {Brown}, \citenamefont {Busing},\ and\
  \citenamefont {Levy}}]{Noespirlet1975}%
  \BibitemOpen
  \bibfield  {author} {\bibinfo {author} {\bibfnamefont {M.~R.}\ \bibnamefont
  {Noe-Spirlet}}, \bibinfo {author} {\bibfnamefont {G.~M.}\ \bibnamefont
  {Brown}}, \bibinfo {author} {\bibfnamefont {W.~R.}\ \bibnamefont {Busing}}, \
  and\ \bibinfo {author} {\bibfnamefont {W.~A.}\ \bibnamefont {Levy}},\
  }\href@noop {} {\bibfield  {journal} {\bibinfo  {journal} {Acta Crystallogr.
  Sec. A}\ }\textbf {\bibinfo {volume} {31}},\ \bibinfo {pages} {Part S3, S80}
  (\bibinfo {year} {1975})}\BibitemShut {NoStop}%
\bibitem [{\citenamefont {Berendsen}\ \emph {et~al.}(1987)\citenamefont
  {Berendsen}, \citenamefont {Grigera},\ and\ \citenamefont
  {Straatsma}}]{Berendsen1987}%
  \BibitemOpen
  \bibfield  {author} {\bibinfo {author} {\bibfnamefont {H.~J.~C.}\
  \bibnamefont {Berendsen}}, \bibinfo {author} {\bibfnamefont {J.~R.}\
  \bibnamefont {Grigera}}, \ and\ \bibinfo {author} {\bibfnamefont {T.~P.}\
  \bibnamefont {Straatsma}},\ }\href {\doibase 10.1021/j100308a038} {\bibfield
  {journal} {\bibinfo  {journal} {J. Phys. Chem.}\ }\textbf {\bibinfo {volume}
  {91}},\ \bibinfo {pages} {6269} (\bibinfo {year} {1987})}\BibitemShut
  {NoStop}%
\bibitem [{\citenamefont {Lee}\ and\ \citenamefont {Rasaiah}(1996)}]{Lee}%
  \BibitemOpen
  \bibfield  {author} {\bibinfo {author} {\bibfnamefont {S.~H.}\ \bibnamefont
  {Lee}}\ and\ \bibinfo {author} {\bibfnamefont {J.~C.}\ \bibnamefont
  {Rasaiah}},\ }\href@noop {} {\bibfield  {journal} {\bibinfo  {journal} {J.
  Phys. Chem.}\ }\textbf {\bibinfo {volume} {100}},\ \bibinfo {pages} {1420}
  (\bibinfo {year} {1996})}\BibitemShut {NoStop}%
\bibitem [{\citenamefont {Koneshan}\ \emph {et~al.}(1998)\citenamefont
  {Koneshan}, \citenamefont {Rasaiah}, \citenamefont {Lynden-Bell},\ and\
  \citenamefont {Lee}}]{Koneshan1998}%
  \BibitemOpen
  \bibfield  {author} {\bibinfo {author} {\bibfnamefont {S.}~\bibnamefont
  {Koneshan}}, \bibinfo {author} {\bibfnamefont {J.~C.}\ \bibnamefont
  {Rasaiah}}, \bibinfo {author} {\bibfnamefont {R.~M.}\ \bibnamefont
  {Lynden-Bell}}, \ and\ \bibinfo {author} {\bibfnamefont {S.~H.}\ \bibnamefont
  {Lee}},\ }\href {\doibase 10.1021/jp980642x} {\bibfield  {journal} {\bibinfo
  {journal} {J. Phys. Chem. B}\ }\textbf {\bibinfo {volume} {102}},\ \bibinfo
  {pages} {4193} (\bibinfo {year} {1998})}\BibitemShut {NoStop}%
\bibitem [{\citenamefont {http://lammps.sandia.gov}()}]{lammps}%
  \BibitemOpen
  \bibfield  {author} {\bibinfo {author} {\bibfnamefont {L.}~\bibnamefont
  {http://lammps.sandia.gov}},\ }\href@noop {} {\ }\BibitemShut {NoStop}%
\bibitem [{\citenamefont {Ryckaert}\ \emph {et~al.}(1977)\citenamefont
  {Ryckaert}, \citenamefont {Ciccotti},\ and\ \citenamefont
  {Berendsen}}]{Ryckaert1977}%
  \BibitemOpen
  \bibfield  {author} {\bibinfo {author} {\bibfnamefont {J.-P.}\ \bibnamefont
  {Ryckaert}}, \bibinfo {author} {\bibfnamefont {G.}~\bibnamefont {Ciccotti}},
  \ and\ \bibinfo {author} {\bibfnamefont {H.~J.}\ \bibnamefont {Berendsen}},\
  }\href {\doibase 10.1016/0021-9991(77)90098-5} {\bibfield  {journal}
  {\bibinfo  {journal} {J. Comput. Phys.}\ }\textbf {\bibinfo {volume} {23}},\
  \bibinfo {pages} {327} (\bibinfo {year} {1977})}\BibitemShut {NoStop}%
\bibitem [{\citenamefont {Martyna}\ \emph {et~al.}(1994)\citenamefont
  {Martyna}, \citenamefont {Tobias},\ and\ \citenamefont
  {Klein}}]{Martyna1994}%
  \BibitemOpen
  \bibfield  {author} {\bibinfo {author} {\bibfnamefont {G.~J.}\ \bibnamefont
  {Martyna}}, \bibinfo {author} {\bibfnamefont {D.~J.}\ \bibnamefont {Tobias}},
  \ and\ \bibinfo {author} {\bibfnamefont {M.~L.}\ \bibnamefont {Klein}},\
  }\href {\doibase 10.1063/1.467468} {\bibfield  {journal} {\bibinfo  {journal}
  {J. Chem. Phys.}\ }\textbf {\bibinfo {volume} {101}},\ \bibinfo {pages}
  {4177} (\bibinfo {year} {1994})}\BibitemShut {NoStop}%
\bibitem [{\citenamefont {Martyna}\ and\ \citenamefont
  {Klein}(1992)}]{Martyna1992}%
  \BibitemOpen
  \bibfield  {author} {\bibinfo {author} {\bibfnamefont {G.~J.}\ \bibnamefont
  {Martyna}}\ and\ \bibinfo {author} {\bibfnamefont {M.~L.}\ \bibnamefont
  {Klein}},\ }\href {\doibase 10.1063/1.463940} {\bibfield  {journal} {\bibinfo
   {journal} {Phys. Rev. Lett.}\ }\textbf {\bibinfo {volume} {2635}},\ \bibinfo
  {pages} {2635} (\bibinfo {year} {1992})}\BibitemShut {NoStop}%
\bibitem [{\citenamefont {Tazi}\ \emph {et~al.}(2012)\citenamefont {Tazi},
  \citenamefont {Botan}, \citenamefont {Salanne}, \citenamefont {Marry},
  \citenamefont {Turq},\ and\ \citenamefont {B.}}]{TaziJPCM2012}%
  \BibitemOpen
  \bibfield  {author} {\bibinfo {author} {\bibfnamefont {S.}~\bibnamefont
  {Tazi}}, \bibinfo {author} {\bibfnamefont {A.}~\bibnamefont {Botan}},
  \bibinfo {author} {\bibfnamefont {M.}~\bibnamefont {Salanne}}, \bibinfo
  {author} {\bibfnamefont {V.}~\bibnamefont {Marry}}, \bibinfo {author}
  {\bibfnamefont {P.}~\bibnamefont {Turq}}, \ and\ \bibinfo {author}
  {\bibfnamefont {R.}~\bibnamefont {B.}},\ }\href@noop {} {\bibfield  {journal}
  {\bibinfo  {journal} {J. Phys.: Condens. Matter}\ }\textbf {\bibinfo {volume}
  {24}},\ \bibinfo {pages} {284117} (\bibinfo {year} {2012})}\BibitemShut
  {NoStop}%
\bibitem [{\citenamefont {Guyon}\ \emph {et~al.}(2001)\citenamefont {Guyon},
  \citenamefont {Hulin},\ and\ \citenamefont {Petit}}]{Guyon2001}%
  \BibitemOpen
  \bibfield  {author} {\bibinfo {author} {\bibfnamefont {E.}~\bibnamefont
  {Guyon}}, \bibinfo {author} {\bibfnamefont {J.-P.}\ \bibnamefont {Hulin}}, \
  and\ \bibinfo {author} {\bibfnamefont {L.}~\bibnamefont {Petit}},\
  }\href@noop {} {\emph {\bibinfo {title} {{Hydrodynamique physique}}}},\
  \bibinfo {edition} {2nd}\ ed.\ (\bibinfo  {publisher} {EDP Sciences, CNRS
  Editions},\ \bibinfo {address} {Paris},\ \bibinfo {year} {2001})\BibitemShut
  {NoStop}%
\bibitem [{\citenamefont {D\"unweg}\ and\ \citenamefont
  {Kremer}(1993)}]{kremer93}%
  \BibitemOpen
  \bibfield  {author} {\bibinfo {author} {\bibfnamefont {B.}~\bibnamefont
  {D\"unweg}}\ and\ \bibinfo {author} {\bibfnamefont {K.}~\bibnamefont
  {Kremer}},\ }\href {\doibase 10.1063/1.465445} {\bibfield  {journal}
  {\bibinfo  {journal} {J. Chem. Phys.}\ }\textbf {\bibinfo {volume} {99}},\
  \bibinfo {pages} {6983} (\bibinfo {year} {1993})}\BibitemShut {NoStop}%
\bibitem [{\citenamefont {Long}\ and\ \citenamefont
  {Cronin}(2006)}]{Cronin2006}%
  \BibitemOpen
  \bibfield  {author} {\bibinfo {author} {\bibfnamefont {D.-L.}\ \bibnamefont
  {Long}}\ and\ \bibinfo {author} {\bibfnamefont {L.}~\bibnamefont {Cronin}},\
  }\href {\doibase 10.1002/chem.200501002} {\bibfield  {journal} {\bibinfo
  {journal} {Chem. Eur. J.}\ }\textbf {\bibinfo {volume} {12}},\ \bibinfo
  {pages} {3698} (\bibinfo {year} {2006})}\BibitemShut {NoStop}%
\bibitem [{\citenamefont {Pusset}\ \emph {et~al.}(2015)\citenamefont {Pusset},
  \citenamefont {Gourdin-Bertin}, \citenamefont {Dubois}, \citenamefont
  {Chevalet}, \citenamefont {M\'{e}riguet}, \citenamefont {Bernard},
  \citenamefont {Dahirel}, \citenamefont {Jardat},\ and\ \citenamefont
  {Jacob}}]{Pusset2015}%
  \BibitemOpen
  \bibfield  {author} {\bibinfo {author} {\bibfnamefont {R.}~\bibnamefont
  {Pusset}}, \bibinfo {author} {\bibfnamefont {S.}~\bibnamefont
  {Gourdin-Bertin}}, \bibinfo {author} {\bibfnamefont {E.}~\bibnamefont
  {Dubois}}, \bibinfo {author} {\bibfnamefont {J.}~\bibnamefont {Chevalet}},
  \bibinfo {author} {\bibfnamefont {G.}~\bibnamefont {M\'{e}riguet}}, \bibinfo
  {author} {\bibfnamefont {O.}~\bibnamefont {Bernard}}, \bibinfo {author}
  {\bibfnamefont {V.}~\bibnamefont {Dahirel}}, \bibinfo {author} {\bibfnamefont
  {M.}~\bibnamefont {Jardat}}, \ and\ \bibinfo {author} {\bibfnamefont
  {D.}~\bibnamefont {Jacob}},\ }\href {\doibase 10.1039/C5CP00487J} {\bibfield
  {journal} {\bibinfo  {journal} {Phys. Chem. Chem. Phys.}\ }\textbf {\bibinfo
  {volume} {17}},\ \bibinfo {pages} {11779} (\bibinfo {year}
  {2015})}\BibitemShut {NoStop}%
\bibitem [{\citenamefont {Laage}\ and\ \citenamefont {Hynes}(2008)}]{Laage}%
  \BibitemOpen
  \bibfield  {author} {\bibinfo {author} {\bibfnamefont {D.}~\bibnamefont
  {Laage}}\ and\ \bibinfo {author} {\bibfnamefont {J.~T.}\ \bibnamefont
  {Hynes}},\ }\href {\doibase 10.1021/jp802033r} {\bibfield  {journal}
  {\bibinfo  {journal} {J. Phys. Chem. B}\ }\textbf {\bibinfo {volume} {112}},\
  \bibinfo {pages} {7697} (\bibinfo {year} {2008})}\BibitemShut {NoStop}%
\bibitem [{\citenamefont {Chowdhuri}\ and\ \citenamefont
  {Chandra}(2001)}]{chowdhuriJCP2001}%
  \BibitemOpen
  \bibfield  {author} {\bibinfo {author} {\bibfnamefont {S.}~\bibnamefont
  {Chowdhuri}}\ and\ \bibinfo {author} {\bibfnamefont {A.}~\bibnamefont
  {Chandra}},\ }\href@noop {} {\bibfield  {journal} {\bibinfo  {journal} {J.
  Chem. Phys.}\ }\textbf {\bibinfo {volume} {115}},\ \bibinfo {pages} {3732}
  (\bibinfo {year} {2001})}\BibitemShut {NoStop}%
\bibitem [{\citenamefont {Zhang}\ \emph {et~al.}(2016)\citenamefont {Zhang},
  \citenamefont {Tran},\ and\ \citenamefont {Gray-Weale}}]{ZhangJPCB2016}%
  \BibitemOpen
  \bibfield  {author} {\bibinfo {author} {\bibfnamefont {X.}~\bibnamefont
  {Zhang}}, \bibinfo {author} {\bibfnamefont {S.}~\bibnamefont {Tran}}, \ and\
  \bibinfo {author} {\bibfnamefont {A.}~\bibnamefont {Gray-Weale}},\
  }\href@noop {} {\bibfield  {journal} {\bibinfo  {journal} {J. Phys. Chem. C}\
  }\textbf {\bibinfo {volume} {120}},\ \bibinfo {pages} {21888} (\bibinfo
  {year} {2016})}\BibitemShut {NoStop}%
\end{thebibliography}

%

\end{document}